\documentclass[preprint,1p,12pt]{elsarticle}

\makeatletter
\def\ps@pprintTitle{%
  \let\@oddhead\@empty
  \let\@evenhead\@empty
  \let\@evenfoot\@oddfoot
}
\makeatother

\usepackage{amssymb,amsmath}
\usepackage[latin1]{inputenc}
\usepackage{color}

\newcommand{\ot}{\leftarrow}
\newcommand{\adjoint}{\mathop{\&}\nolimits}

\renewcommand{\Dot}{\hbox{\boldmath $.$}}
\newcommand{\Dotted}[1]{\mathop{\stackrel\Dot{#1}}\nolimits}

\def\P{\ensuremath{\mathbb{P}}}

\newtheorem{theorem}{Theorem}

\newtheorem{lemma}[theorem]{Lemma}
\newtheorem{proposition}[theorem]{Proposition}
\newdefinition{definition}[theorem]{Definition }
\newdefinition{remark}[theorem]{Remark }
\newdefinition{example}[theorem]{Example }
\newproof{proof}{Proof}

\begin{document}

\pagenumbering{arabic}

\begin{frontmatter}

\title{Syntax and   semantics of \\multi-adjoint normal logic programming}

\author
{M. Eugenia Cornejo, David Lobo, Jesús Medina}

\address
{Department of Mathematics,
 University of  C\'adiz. Spain\\
\texttt{\{mariaeugenia.cornejo,david.lobo,jesus.medina\}@uca.es}}

\begin{abstract}
Multi-adjoint logic programming is a general framework with interesting features, which involves other positive logic programming frameworks such as
monotonic and residuated logic programming, generalized annotated logic programs, fuzzy logic programming and  possibilistic logic programming.

 One of the most interesting extensions of this framework is the possibility of   considering   a negation operator in the logic programs, which will improve its  flexibility and the range of real applications.

This paper introduces multi-adjoint normal logic programming, which is an extension of multi-adjoint logic programming   including a negation operator in the underlying lattice. Beside the introduction of the syntax and semantics of this paradigm, we will      provide sufficient conditions for the existence of stable models  defined on  a convex compact set of an euclidean space. 
Finally, 
we will consider a particular   algebraic structure in which  sufficient conditions can be given in order to ensure the unicity of stable models of multi-adjoint normal logic programs.

\end{abstract}


\end{frontmatter}

\section{Introduction}

Syntax and semantics are two noticeably different parts in any theory of logic programming. On the one hand, the syntax describes the symbols and formulas chosen to represent formally statements that can be considered. On the other hand, the semantics gives meaning to the considered statements from its syntactic structure, and establishes an inference system to obtain which deductions and/or consequences are correct. In this paper, we will regard a specific type of   structure called multi-adjoint logic programs.
Multi-adjoint logic programming was introduced as a generalization of different non-classical logic programming frameworks in~\cite{lpnmr01}. The main feature of this logical theory is based on the use of several implications in the rules of a same logic program, as well as general operators defined on complete lattices in the bodies of the rules. An interesting property of multi-adjoint logic programs is related to the existence of the least model. This fact allows us to check whether a statement is a consequence by using a simple evaluation. However, the existence of the least model cannot be guaranteed when we consider  multi-adjoint logic programs enriched with a negation operator.
Different semantics have been developed for logic programs with negation such as the well-founded semantics~\cite{acm1991}, the stable models semantics~\cite{iclp88} and the answer sets semantics~\cite{ngc1991}. This work will focus on the study of the existence and the unicity of stable models for multi-adjoint normal logic programs. In some logical approaches, sufficient conditions have been stated in order to ensure the existence of stable models:
\begin{itemize}
	\item In the 3-valued Kleene logic, every logic program with negation has stable models~\cite{Przymusinski}.
	\item In normal residuated logic programming, stable models exist for every normal logic program  whose underlying residuated lattice has an appropriate bilattice structure~\cite{Fitting,Loyer,Straccia1,Straccia2,Straccia3}.
	\item Another important survey on stable models in normal residuated logic programs on $[0,1]$ was presented in~\cite{Madrid}, where it was proven that the continuity of the connectives appearing in the program guarantees the existence of stable models. In addition, the uniqueness of stable models is obtained when the product t-norm, its residuated implication, and the standard negation are considered.
	\end{itemize}

The contribution of this paper   consists in applying the philosophy of the multi-adjoint paradigm~\cite{escim2016,lpnmr01} in order to develop a more general and flexible mathematical theory than the previous ones. Hence, we will present the syntax and semantics of the  multi-adjoint normal logic programming framework  and   sufficient conditions to ensure:

\begin{enumerate}
	\item the existence of stable models for multi-adjoint  normal logic programs  defined on any convex compact set of an euclidean space; and
	\item the unicity of stable models for multi-adjoint normal logic programs defined on the set of subintervals  $\mathcal{C}([0,1])$.
\end{enumerate}

When these logic programs correspond to some search problem, the stable models coincide with its possible solutions. Therefore, these goals are fundamental in order to know whether the program is related to a solvable problem and, in that case, whether only one solution exists. The characterization of programs with a unique solution and a deterministic procedure to obtain that solution is also important, since the solvability of this kind of programs will be at once.

These properties on the existence and uniqueness of stable models in multi-adjoint  normal logic programming will be useful in other logic programming frameworks in which a  negation operator is needed. Since monotonic and residuated logic programming~\cite{DP01:ecsqaru,DP-jelia}, fuzzy logic programming~\cite{vojtas-fss} and possibilistic logic programming~\cite{dlp:1994} are particular cases of the multi-adjoint  logic programming  framework,  we can straightforwardly  apply the results given in this paper  to their normal extensions (that is, when a negation operator is used).

 Moreover, these results can be also applied to other frameworks with a different syntax, such as to  generalized annotated logic programs~\cite{gaps}. This logic  was related to the fuzzy logic programming introduced by Vojtás in~\cite{klv}, and so we can  translate the results given in this paper to  a new normal generalized annotated logic programming  in which a  negation operator is considered.

This paper is organized as follows: Section~\ref{sec:preliminares} includes a brief summary with concepts and results corresponding to the multi-adjoint  logic programming framework and the algebraic topology. Section~\ref{sec:sinandsem} presents  the multi-adjoint normal logic programs as well as interesting properties of the immediate consequence operator and of the stable models. These properties allow to recognize which are the problems we have to solve in order to define the syntax and the semantics of multi-adjoint normal logic programs. A detailed study about the existence and the unicity of stable models in these programs is introduced in Section~\ref{sec:exi-uni}. Some conclusions and prospects for future work are included in Section~\ref{sec:conclusion}.

\section{Preliminaries}\label{sec:preliminares}

This section   recalls some notions and results related to the propositional language used in multi-adjoint logic programming, which is composed of two important parts: the syntax and the semantics. Later, the definition of program in this general logic programming framework is included. Finally, some topological definitions and fix-point theorems will be introduced. 

\subsection{Syntax of the propositional language}
The syntax of the propositional language of multi-adjoint logic programming is based on the concepts of alphabet and expressions of the language. These concepts require the use of some definitions of universal algebra as it is shown below.

\begin{definition} \label{def:gradedset}
A \emph{graded set} is a set $\Omega$ with a function which assigns to each element $\omega \in \Omega$ a number $n\geq 0$ called the arity of $\omega$. The set $\Omega_n$ will denote the set of elements with arity $n$ in $\Omega$.
\end{definition}

Considering a graded set, the notions of algebraic structure and substructure of an algebraic structure are generalized by means of the following definitions.

\begin{definition} \label{def:sigmaalgebra}
Given a graded set $\Omega$, an {\em $\Omega$-algebra} is a pair $\mathfrak{A}=\langle A,I\rangle$ where $A$ is a non-empty set
called the carrier, and $I$ is a function which assigns maps to the elements of $\Omega$ as follows:
\begin{enumerate}
\item Each element  $\omega \in \Omega_n$, $n > 0$,  is interpreted as a map $I(\omega) \colon A^n \to A$, denoted by
$\omega_\mathfrak{ A}$.
\item Each element $c \in \Omega_0$  ($c$ is a constant) is interpreted as an element $I(c)$ in $A$, denoted by $c_\mathfrak{
A}$.
\end{enumerate}
\end{definition}

\begin{definition}\label{def:subalgebra}
 Given an $\Omega$-algebra  $\mathfrak{ A}=\langle A,I\rangle$, an {\em $\Omega$-subalgebra $\mathfrak{ B}$} is a pair
$\langle B,J\rangle$, such that $B\subseteq A$ and
\begin{enumerate}
\item $J(c) = I(c) $ for all $c\in \Omega_0$.
\item Given $\omega\in \Omega_n$, then $J(\omega)\colon B^n \to B$ is
the restriction of $I(\omega)\colon A^n\to A$.
\end{enumerate}
\end{definition}

Now, we introduce the notion of alphabet of a language, that is, the set of symbols from which expressions can be formed.

\begin{definition}\label{def:alphabet}
Let $\Omega$  be a graded set, $\Pi$ a countable infinite set and $L$ a set of truth-values. The {\em alphabet ${ A}_{\Omega,\Pi\uplus L}$}  associated with $\Omega$ and $\Pi\uplus L$ is defined by the disjoint union $\Omega \uplus( \Pi \uplus L)\uplus S $, where $S$ is the set of auxiliary symbols  ``('', ``)'' and ``,''.
\end{definition}

From the set of operators $\Omega$ and the symbols of $\Pi\uplus L$, the algebra of expressions is defined as follows.

\begin{definition}\label{def:expressions}
Given a graded set $\Omega$ and an alphabet
   ${ A}_{\Omega,\Pi\uplus L}$. The $\Omega$-algebra $\mathfrak{ E}=\langle
{ A}_{\Omega,\Pi\uplus L}^*,I\rangle$ of \emph{expressions} is defined as follows:
\begin{enumerate}
\item The carrier ${ A}_{\Omega,\Pi\uplus L}^*$ is the set of strings over ${ A}_{\Omega,\Pi\uplus L}$.
\item The interpretation function $I$ satisfies the following conditions for
      strings $a_1, \dots, a_n $ in~${ A}_{\Omega,\Pi\uplus L}^*$:
\begin{itemize}
\item  $c_\mathfrak{ E} = c$, where $c$ is a constant operation ($c\in \Omega_0$).
\item  $\omega_\mathfrak{ E}(a_1) = \omega \, a_1$, where $\omega$ is
an unary operation  ($\omega \in \Omega_1$).
\item  $\omega_\mathfrak{ E}(a_1,a_2) = (a_1\omega \, a_2)$, where
$\omega$ is a binary operation  ($\omega \in \Omega_2$).
\item  $\omega_\mathfrak{ E}(a_1,\ldots,a_n)  =
\omega(a_1,\dots,a_n)$, where $\omega$ is a n-ary operation ($\omega\in \Omega_n$) and $n>2$.
\end{itemize}
\end{enumerate}
\end{definition}

It is important to note that an expression does not need to be a well-formed formula, that is, an expression is only a string of letters of the alphabet. Indeed, the well-formed formulas is the subset of expressions defined as the next definition shows.

\begin{definition}\label{def: well-formed formulas}
Let  $\Omega$ be a  graded set, $\Pi$  a countable set of propositional symbols, $L$ a set of truth-values and $\mathfrak E$ the  algebra of expressions corresponding to the alphabet $A_{\Omega,\Pi\uplus L}$. The \emph{well-formed formulas} (in short, formulas) generated by $\Omega$ over $\Pi\uplus L$  is the least subalgebra $\mathfrak{F}$ of the algebra of expressions $\mathfrak{ E}$  containing $\Pi\uplus L$.
\end{definition}

\subsection{Semantics of the propositional language}\label{sec:semlanguage}

In this section, we will consider a graded set $\Omega$, a set of propositional symbols $\Pi$, the corresponding $\Omega$-algebra of well-formed formulas $\mathfrak{F}$ and an arbitrary $\Omega$-algebra $\mathfrak{U}$ whose carrier is $A$. The notion of interpretation plays a fundamental role in the semantics of the propositional language of multi-adjoint logic programming.

\begin{definition}
A  mapping $I\colon\Pi\to A$ which assigns to every propositional symbol appearing in $\Pi$  an element of $A$ is called \emph{$A$-interpretation}. The set of all $A$-interpretations with respect to the $\Omega$-algebra  $\mathfrak U$ is denoted by $\cal{I}_\mathfrak{U}$.
\end{definition}

If $(L,\preceq)$ is a complete lattice where $L$ is the carrier of an $\Omega$-algebra $\mathfrak L$ then the ordering $\preceq$ can be extended to the set of interpretations as follows:
$$ I_1 \sqsubseteq I_2 \hbox{ if and only if } I_1(p) \preceq I_2(p) \text{, for all } p \in \Pi \text{ and } I_1,I_2 \in\cal{I}_\mathfrak{L}\text{.}$$ 

The new ordering $\sqsubseteq$ defined on the set of interpretations inherits some properties of the ordering $\preceq$ defined on the lattice, as the next proposition shows.

\begin{proposition}[\cite{lpnmr01}]\label{prop:set:inter}
If $(L,\preceq)$ is a complete lattice, then $(\cal{I}_\mathfrak{L},\sqsubseteq)$ is a complete lattice where the least interpretation $\Delta$ applies every propositional symbol to the bottom element of $L$, and the greatest interpretation $\nabla$ applies every propositional symbol to the top element of $L$.
\end{proposition}

A similar result with respect to the convexity and the compactness of the set of interpretations will be proved in   Section~\ref{sec:exstable}, which is  focused on the study of the existence of stable models. 

\subsection{Multi-adjoint logic programs}
The multi-adjoint framework arises as a generalization of several non-classical logic programming settings whose
semantic structure is the multi-adjoint lattice~\cite{fss:cmr:2013,ija-cmr15,lpnmr01}. In order to recall this definition, we need to introduce the concept of adjoint pair which was firstly presented in a logical context by Pavelka~\cite{pavelka}.

\begin{definition} \label{def:adjointpair}
Let $( P,\leq) $ be a partially ordered set and $(\adjoint{},\ot)$ be a pair of binary operations in $P$, such that
\begin{enumerate}
\item $\adjoint$ is monotonic in both arguments.\footnote{A monotonic operator is also called  order-preserving operator or increasing mapping.}
\item $\ot$ is monotonic in the first  argument (the consequent) and decreasing in the second argument (the antecedent).
\item For  any $x,y,z \in P$\@, we have that $x \leq  (y \ot z)$ holds  if and only if $(x \adjoint{} z) \leq  y$ holds.
\end{enumerate}
Then we say that $(\adjoint{},\ot)$ forms an \emph{adjoint pair} in $(P,\leq ) $.
\end{definition}

Observe that, the monotonicity of the operators $\adjoint$ and $\ot$ is justified because they will be interpreted as generalized conjunctions and implications. It is important to highlight that $\adjoint$ does not need to be either commutative or associative, and boundary conditions are not required. The last property in the previous definition corresponds to the categorical adjointness.

As well as the properties given in Definition~\ref{def:adjointpair}, we will need to assume the existence of the bottom and top elements in the poset of truth-values, and the existence of joins for every directed subset, that is, we will assume a complete lattice.

The use of different implications and several modus ponens like inference rules to extend the theory developed in~\cite{DP01:ecsqaru,vojtas-fss} to a more general environment gave rise to consider various adjoint pairs in the lattice.

\begin{definition} \label{multi-adjoint}
The tuple $(L,\preceq,\leftarrow_1,\adjoint_1,\dots,\leftarrow_n,\adjoint_n )$ is a \emph{multi-adjoint lattice} if the following properties are verified:
\begin{enumerate}
\item  $( L,\preceq ) $ is a bounded lattice, i.e.\ it has a bottom $(\bot)$ and a top $(\top)$ element;
\item $(\adjoint_i,\ot_i)$ is an adjoint pair in $( L,\preceq ) $, for $i\in\{1, \dots,n\}$;
\item  $\top \adjoint_i \vartheta = \vartheta \adjoint_i \top=\vartheta$, for all $\vartheta\in L$ and $i\in\{1, \dots, n\}$.
\end{enumerate}
\end{definition}

The algebraic structure shown in the next definition increases the expressive power of the multi-adjoint lattice by using extra operators.

\begin{definition} \label{multi-adjoint algebra}
Let $\Omega$ be a graded set  containing operators  $\leftarrow_i$ and $\adjoint_i$ for $i\in\{1, \dots, n\}$  and possibly some extra operators, and let $\mathfrak{ L}=(L,I)$ be an  $\Omega$-algebra  whose carrier set $L$ is a lattice under~$\preceq$. We say  that  $\mathfrak{ L}$   is a \emph{multi-adjoint $\Omega$-algebra} with respect to the pairs $(\adjoint_i,\ot_i)$, with $i\in\{1, \dots, n\}$, if
$(L,\preceq,I({\leftarrow_1}),I({\adjoint_1}), \dots, I({\leftarrow_n}),I({\adjoint_n}))$ is a multi-adjoint lattice.
\end{definition}

From this structure, a multi-adjoint logic program is defined as a set of rules and facts of a given language $\mathfrak  F$.

\begin{definition}\label{def:malp}  Let $(L,\preceq,\leftarrow_1,\adjoint_1,\dots,\leftarrow_n,\adjoint_n )$ be a multi-adjoint lattice. A
\emph{multi-adjoint logic program} is a set of weighted rules \(\left\langle( A \leftarrow_i  \mathcal{B} ); \vartheta
\right\rangle \) such that:
\begin{enumerate}
\item The \emph{rule} $( A \leftarrow_i \mathcal{B} )$ is a formula of $\mathfrak  F$\@;
\item The \emph{confidence factor} $\vartheta$ is an element (a truth-value) of $ L$;
\item The \emph{head}  of the rule $A$ is a propositional symbol of $\Pi$\@.
\item The \emph{body}  formula $\mathcal{B}$ is a formula of $\mathfrak F$ built from propositional  symbols $B_1,\ldots,B_n$ ($n \geq 0$) by the use of conjunctors $\adjoint_1,\dots,\adjoint_{n}$ and $\land_1,\dots, \land_k$, disjunctors $\lor_1,\dots,\lor_l$, aggregators $@_1,\dots,@_m$ and elements of $L$.
\item \emph{Facts} are  rules with body $ \top$.
\end{enumerate}
\end{definition}

Examples related to these preliminary notions can be found in~\cite{fss-gines,jal,iclp01,lpnmr01}.

Note that, when the multi-adjoint lattice is enriched with a negation operator, we can define a particular type of multi-adjoint logic program called multi-adjoint normal logic program. Before presenting our study about the syntax and semantics of this special kind of non-monotonic logic program, we need to recall some topological notions and results.

\subsection{Some notions of algebraic topology}

This section includes different notions and results of algebraic topology, which will be used later. 
The definitions of compact set and convex set are listed below.

\begin{definition}
Let $(X,+,*,\mathbb{R})$ be an euclidean space. We say that $A\subseteq X$ is:
\begin{itemize}
\item a \emph{compact set} if it is closed and bounded in $X$.
\item a \emph{convex set} if $t* x+(1-t)*y\in A$, for all $x,y\in A$ and $t\in[0,1]$.
\end{itemize}
\end{definition}

Finally, we present two theorems related to the fix-point theory. The former is an extension of the Brouwer fix-point theorem, which is known as Schauder fix-point theorem~\cite{Leborgne}.

\begin{theorem}[Schauder fix-point theorem]\label{Schauder}
    Let $(X,+,*,\mathbb{R})$ be an euclidean space and let $K\subseteq X$ be a non-empty convex compact set. Every continuous mapping $f\colon K \to K$ has a fix point.
\end{theorem}

Before introducing the Banach fix-point theorem~\cite{Banach1922}, it is necessary to show the definition of contractive mapping.
\begin{definition}
 Let $(X,d)$ be a complete metric space. We say that $f\colon  X\rightarrow X$ is \emph{a contractive mapping} if there  exists a real value $0<\lambda<1$ such that:
$$
d(f(x),f(y))\leq \lambda\, d(x,y)
$$
for each $x,y\in X$. Any real value $\lambda$ satisfying the previous inequality is called \emph{Lipstchiz constant}.
\end{definition}

\begin{theorem}[Banach fix-point theorem]\label{Banach}
    Let $(X,d)$ be a complete metric space and let $f\colon  X\rightarrow X$ be a contractive mapping in $A\subseteq X$. Then $f$ has a unique fix-point in $A$.
\end{theorem}

These previous concepts and results will play a crucial role in order to define the semantics of multi-adjoint normal logic programs.

\section{On the syntax and semantics of multi-adjoint normal logic programs}\label{sec:sinandsem}
As we mentioned above, we are interested in considering a multi-adjoint lattice  $(L,\preceq,\leftarrow_1,\adjoint_1,\dots,\leftarrow_n,\adjoint_n )$, with a maximum $\top$ and  a minimum $\bot$ element, enriched with  a negation operator. The considered negation will be a decreasing mapping $\neg\colon L\rightarrow L$ satisfying the equalities $\neg(\bot)=\top$ and $\neg(\top)=\bot$. The notion of default negation is modeled by the previous negation operator. The algebraic structure obtained from a multi-adjoint lattice and a negation operator will be called a \emph{multi-adjoint normal lattice}.

The formal definition of a multi-adjoint normal logic program is given next.

\begin{definition}\label{def:nmalp}
  A \emph{multi-adjoint normal logic program (MANLP)} $\P$, defined on  a multi-adjoint normal lattice $(L,\preceq,\leftarrow_1,\adjoint_1,\dots,\leftarrow_n,\adjoint_n,\neg )$, is a finite set of weighted rules of the form:
$$\langle p\leftarrow_i @[ p_1, \dots, p_m,\neg p_{m+1},\dots,\neg p_n]; \vartheta \rangle$$
where $i\in\{1,\dots,n\}$, $@$ is an aggregator operator, $\vartheta$ is an element of $L$ and $p,p_1,\dots,p_n$ are propositional symbols such that $p_j\neq p_k$, for all $j,k\in\{1,\dots,n\}$, with $j\neq k$.
\end{definition}
	
Henceforth, we will use the notation $\Pi_\P$ to denote the set of propositional symbols appearing in $\P$. In addition, the rules of a MANLP will be denoted as $\langle p\leftarrow_i \mathcal{B};\vartheta \rangle$ where $p$ is the \emph{head} of  the rule, $\mathcal{B}$ its \emph{body} and $\vartheta$ its \emph{weight}.

It is also convenient to mention that the whole set of rules that we can build from the well-formed formulas generated by $\Omega$ over $\Pi\uplus L$ will be denoted by $\mathfrak{R}_{\Pi\uplus L}$. The set $\mathfrak{R}^+_{\Pi\uplus K}$ will be formed by  the  rules  of $\mathfrak{R}_{\Pi\uplus K}$ which do not contain the  negation operator.

In order to avoid confusion, we will use a special notation to differentiate   an operator symbol in $\Omega$ from  its interpretation under $\mathfrak{L}$. Specifically, $\omega$ will denote an operator symbol in $\Omega$ and $\Dotted \omega$ will denote the interpretation of the previous operator symbol under $\mathfrak{L}$. In a similar way, the evaluation of a formula $\cal A$ under an interpretation $I$ will be denoted as $\hat{I}({\cal A})$, and it proceeds inductively as usual, until all propositional symbols in ${\cal A}$ are reached and evaluated under $I$. For instance, considering an interpretation $I \in \cal{I}_\mathfrak{L}$ and two formulas $\cal A, \cal B\in\mathfrak  F$, the following equality holds:
\[
\hat{I}( { \cal A} \adjoint_{i} {\cal B })=\hat{I}({\cal A})  \Dotted {\adjoint}_{i}
\hat{I} ({\cal B})
\]

Note that, every formula $\cal A$  can be written as $@[p_1, \dots, p_m,\neg p_{m+1},\dots,\neg p_n]$, in which $@$ represents the composition of the monotonic operators in $\cal A$ (which is an aggregation operator) and $p_1,\dots,p_n$ the  propositional symbols appearing in $\cal A$. In this case, the equality $\hat{I} ( @[ p_1, \dots, p_m,\neg p_{m+1},\dots,\neg p_n])=\Dotted{@}[{I} (p_1), \dots, I( p_m),\neg I(p_{m+1}),\dots,\neg {I} (p_n)]$ is satisfied for any formula ${\cal A}= @[ p_1, \dots, p_m,\neg p_{m+1},\dots,\neg p_n]\in \mathfrak  F$.

Now, after introducing the  syntactic structure of MANLPs and some notational conventions, we will present the notions associated with the semantics of MANLPs. We will start with the definitions of satisfaction and model.
	
In a similar way to the semantics of multi-adjoint logic programs~\cite{lpnmr01}, we say that an interpretation satisfies a rule of a multi-adjoint normal logic program if the truth-value of the rule is greater or equal than the confidence factor associated with the rule.

\begin{definition} Given an interpretation $I \in \cal{I}_\mathfrak{L}$\@, we say that:
\begin{itemize}
\item[(1)] A weighted  rule $\langle p\leftarrow_i @[ p_1, \dots, p_m,\neg p_{m+1},\dots,\neg p_n]; \vartheta \rangle$
is \emph{satisfied} by $I$ if and only if  $\vartheta\preceq \hat{I}\left( p\leftarrow_i @[ p_1, \dots, p_m,\neg p_{m+1},\dots,\neg p_n]
\right)$.
\item[(2)]An interpretation $I \in \cal{I}_\mathfrak{L}$ is a \emph{model} of a MANLP $\mathbb{P}$ if and only if all
weighted rules in $\mathbb{P}$ are  satisfied by~$I$\@.
\end{itemize}
\end{definition}

\begin{example}\label{ex:preliminar}
	Consider the multi-adjoint normal lattice
	$$\langle[0,1],\leq,\leftarrow_G,\adjoint_G,\leftarrow_P,\adjoint_P,\neg\rangle$$
	where $\adjoint_G$ and $\adjoint_P$ are the Gödel and product conjunctors, respectively, $\leftarrow_G$ and $\leftarrow_P$ are their corresponding adjoint implications and $\neg(x)=1-x$, for each $x\in[0,1]$.
	
	Let $\Pi_\P=\{p,q,r\}$ be the set of propositional symbols and let us define the following MANLP $\P$ valued in $[0,1]$ and consisting of two rules and one fact.
	\begin{equation*}
	\begin{array}{l}
	r_1:\ \langle p\leftarrow_P q\ \adjoint_G\neg r\ ;\ 0.7\rangle\\
	r_2:\ \langle r\leftarrow_G p\adjoint_G q\ ;\ 0.2\rangle\\
	r_3:\ \langle q\leftarrow_P 1\ ;\ 0.6\rangle
	\end{array}
	\end{equation*}
	
	Let us prove that the interpretation $I\equiv\{(p,0.5),(q,0.7),(r,0.4)\}$ satisfies the rules of $\P$.
	
	For $r_1$ we obtain that
	\begin{eqnarray*}
	\hat{I}(p\leftarrow_P q\ \adjoint_G\neg r)&=&I(p)\Dotted\leftarrow_P\hat{I}\big(q\ \adjoint_G\neg r\big)=I(p)\Dotted\leftarrow_P\big(I(q)\Dotted\adjoint_G\Dotted\neg I(r)\big)\\
	&=&0.5\Dotted\leftarrow_P\big(0.7\Dotted\adjoint_G\Dotted\neg 0.4\big)=0.5\Dotted\leftarrow_P 0.6=\frac{0.5}{0.6}=0.8\widehat{3}
	\end{eqnarray*}
	Since the truth-value of $r_1$ is $0.7$, we have that $0.7\leq \hat{I}(p\leftarrow_P\ q\ \adjoint_G\neg r)$, so $I$ satisfies rule $r_1$.
	
	Considering  rule $r_2$ we obtain that
	\begin{eqnarray*}
	\hat{I}(r\leftarrow_G \ p\adjoint_G\ q)&=&I(r)\Dotted\leftarrow_G\hat{I}\big(p\adjoint_G\ q\big)=I(r)\Dotted\leftarrow_G\big(I(p)\Dotted\adjoint_G I(q)\big)\\
	&=&0.4\Dotted\leftarrow_G\big(0.5\Dotted\adjoint_G 0.7\big)=0.4\Dotted\leftarrow_G 0.5=0.4
	\end{eqnarray*}
	As the weight of rule $r_2$ is $0.2$, we have that $0.2\leq \hat{I}(r\leftarrow_G \ p\adjoint_G\ q)$, hence $I$ satisfies $r_2$.
	
	Lastly, observe that rule $r_3$ is a fact, then $I$ satisfies $r_3$ if and only if $I(q)$ is greater or equal than the weight of $r_3$, which holds, since $I(q)=0.7$ and the weight of $r_3$ is $0.6$. Therefore, $I$ satisfies the three rules in $\P$ and we can conclude that it is a model of that program.\qed
\end{example}

The following section introduces the  immediate consequence operator and the fix-point semantics of MANLPs.

\subsection{Immediate consequence operator}

The first definition generalizes the usual notion of immediate consequence operator for the flexible  case of  multi-adjoint normal logic programs.

\begin{definition}
	Let $\P$ be a multi-adjoint normal logic program. The \emph{immediate consequence operator} is the mapping $T^\mathfrak{L}_\mathbb{P}\colon{\cal I}_\mathfrak{L}\rightarrow {\cal I}_\mathfrak{L}$ defined for every $L$-interpretation $I$ and $p\in\Pi_\P$ as
$$
T_\P(I)(p)=\sup\{\vartheta\Dotted{\adjoint_i} \hat{I}(\mathcal{B}) \mid \langle p\leftarrow_i \mathcal{B};\vartheta \rangle\in\P \}
$$
\end{definition}

The following proposition ensures that, given a MANLP, we can obtain a partition of that program such that, for each propositional symbol $p$, there exists  at most one rule in $\P$ whose head is $p$ in each element of the partition. The interest of this result is that each part of the partition could be considered as an independent program. Note that, if for each propositional symbol $p$ in a program $\P$, there exists at most one rule in $\P$ whose head is $p$, then the definition of $T_\P$ can be simplified, since the supremum operator can be removed from the definition.

\begin{proposition} \label{prop:2.8NicoyManolo}
	Given a MANLP $\P$, there exists a partition $\{\P_\gamma\}_{\gamma\in \Gamma}$ of the program $\P$ such that:
	\begin{enumerate}
		\item $\P_\gamma$ does not contain two rules with the same head, for all $\gamma\in \Gamma$.
		\item the equality $T_\P(I)(p)=\sup\{T_{\P_\gamma}(I)(p)\mid\gamma\in \Gamma\}$ holds.
	\end{enumerate}
\end{proposition}
	\begin{proof}		
		For each rule $r_\gamma\in\P$, let us consider the MANLP with only one rule $\P_\gamma=\{r_\gamma\}$. Then, the partition $\{\P_\gamma\}_{\gamma\in \Gamma}$ satisfies the first condition. Now, for each $\P_\gamma$ and interpretation $I$, the immediate consequence operator is:		
		\begin{equation*}
		T_{\P_\gamma}(I)(q)=\left\{
		\begin{array}{ll}
		\vartheta \Dotted\adjoint_i \hat{I}(\mathcal{B}) & \text{if }\ q=p\\
		\bot & \text{otherwise}
		\end{array}
		\right.
		\end{equation*}
				\noindent
		where $\langle p\leftarrow_i \mathcal{B};\vartheta \rangle$ is the unique rule in $\P_\gamma$. Thus,
		$$T_\P(I)(p)=\sup\{\vartheta \Dotted\adjoint_i \hat{I}(\mathcal{B})\mid\langle p\leftarrow_i \mathcal{B};\vartheta \rangle\in\P\}=\sup\{T_{\P_\gamma}(I)(p)\mid\gamma\in \Gamma\}$$
		\qed
	\end{proof}

This proposition will play a crucial role in the proof of the unicity result of Section~\ref{sec:exi-uni}.

Another important property is that, if  $\P$ is a \emph{positive multi-adjoint logic program}, that is the rules in $\P$ do not contain any negation, we can ensure that its corresponding immediate consequence operator $T_\P$ is monotonic.

\begin{proposition}[\cite{lpnmr01}]\label{monot:ico}
If $\P$ is a positive multi-adjoint logic program, then $T_\P$ is monotonic.
\end{proposition}

This fact allows to characterize the models of a positive multi-adjoint logic program by means of the postfix-points of $T_\P$.

\begin{proposition}[\cite{lpnmr01}]\label{prop6}
Let $\P$ be a positive multi-adjoint logic program and $M$ be an $L$-interpretation. Then, $M$ is a model of the program $\P$ if and only if $T_\P(M)\preceq M$.
\end{proposition}

Note that, when $\P$ is positive, then the Knaster-Tarski fix-point theorem~\cite{tarski1955} ensures that $T_\P$ has a least fix-point. As a consequence, considering the monotonicity of $T_\P$ and the proposition above, we deduce that this least fix-point is the least model of $\P$~\cite{lpnmr01}.

However, the immediate consequence operator is not necessarily monotonic in MANLPs. This fact implies that the existence of the least model cannot be ensured. In order to define the semantics for multi-adjoint normal logic programs, we will use the well-known  notion of stable model of a program~\cite{iclp88,NMadrid1}.

\subsection{Stable models}
The notion of stable model of a normal program is related to the minimal models of a monotonic logic program obtained from the original one. Hence, first of all, we need to introduce a mechanism in order to obtain a positive multi-adjoint logic program from a MANLP.

Given a multi-adjoint normal logic program $\P$ and an $L$-interpretation $I$, we will build a positive multi-adjoint program $\P_I$ by substituting each rule in $\P$ such as
$$\langle p\leftarrow_i @[ p_1, \dots, p_m,\neg p_{m+1},\dots,\neg p_n]; \vartheta \rangle$$
by the rule
$$\langle p\leftarrow_i @_{I}[ p_1, \dots, p_m]; \vartheta \rangle$$
where the operator $\Dotted@_{I}\colon L^m \rightarrow L$ is defined as
$$
\Dotted@_{I}[ \vartheta_1, \dots, \vartheta_m]\!=\!\Dotted @[\vartheta_1,\dots,\vartheta_m,\Dotted \neg {I}(p_{m+1}),\dots, \Dotted \neg{I}(p_n)]
$$
for all $\vartheta_1,\dots,\vartheta_m\in L$. The program $\P_I$ will be called the \emph{reduct} of $\P$ with respect to the interpretation $I$ and  the rules of  the program $\P_I$ will be denoted as $\langle p\leftarrow_i \mathcal{B_I};\vartheta \rangle$.

Then, we say that $I$ is a stable model of $\P$ if and only if $I$ is a minimal model of the reduct $\P_I$.

Now, we can present the definition of stable model of a MANLP.

\begin{definition}\label{def:stable}
Given  a MANLP $\P$ and an  $L$-interpretation $I$, we say that $I$ is a \emph{stable model} of $\P$ if and only if $I$ is a minimal model of $\P_I$.
\end{definition}

Indeed, each stable model of a MANLP $\P$ is a minimal model of $\P$ as the next result shows.

\begin{proposition}\label{prop:2.11NicoyManolo}
Any stable model of a MANLP $\P$ is a minimal model of~$\P$.
\end{proposition}
\begin{proof}
	Let $I$ be a stable model of $\P$. By definition, $I$ is a minimal model of the program $\P_I$. We will prove by reductio ad absurdum that $I$ is a minimal model of $\P$.
	
	Suppose that there exists an interpretation $J$ such that it is a model of $\P$ and $J\sqsubset I$. That is, $J(p)\prec I(p)$ for each $p\in\Pi_\P$. If we prove that $J$ is a model of $\P_I$, we will obtain a contradiction, since $I$ is the minimal model of $\P_I$.
	
	As $J$ is a model of $\P$, for each rule in $\P$ of the form
	$$\langle p\leftarrow_i @[ p_1, \dots, p_m,\neg p_{m+1},\dots,\neg p_n]; \vartheta \rangle$$
	we obtain that
	$$\vartheta \preceq \hat{J}(p\leftarrow_i @[ p_1, \dots, p_m,\neg p_{m+1},\dots,\neg p_n])$$
	That is,
	$$
	\vartheta \preceq J(p)\Dotted\leftarrow_i \Dotted @[J(p_1),\dots,J(p_m),\Dotted \neg  J(p_{m+1}),\dots, \Dotted \neg J(p_n)]
	$$
	Because the operator $\neg$ is decreasing, we obtain that $\Dotted \neg I(p_k)\prec\Dotted \neg J(p_k)$ for all $k\in\{m+1,\dots,n\}$. Hence, since $\Dotted @$ is monotonic, we can ensure that
	$$
	{\footnotesize \Dotted @[J(p_1),\dots,J(p_m),\!\Dotted \neg  \!I(p_{m+1}\!),\dots, \!\Dotted \neg\! I(p_n)]\prec\Dotted @[J(p_1),\dots,J(p_m),,\!\Dotted \neg  \!J(p_{m+1}\!),\dots, \!\Dotted \neg\! J(p_n)]}
	$$
	Finally, as the operator $\leftarrow_i$ is decreasing in the antecedent, we can conclude that
	\begin{eqnarray*}
		\vartheta&\preceq&J(p)\Dotted\leftarrow_i \Dotted @[J(p_1),\dots,J(p_m),\Dotted \neg  J(p_{m+1}),\dots, \Dotted \neg J(p_n)]\\
		&\preceq&J(p)\Dotted\leftarrow_i \Dotted @[J(p_1),\dots,J(p_m),\Dotted \neg  I(p_{m+1}),\dots, \Dotted \neg I(p_n)]\\
		&=&J(p)\Dotted\leftarrow_i\Dotted@_{I}[ J(p_1),\dots,J(p_m)]
	\end{eqnarray*}
	so $J$ is a model of $\P_I$, which contradicts the hypothesis.
	\qed
\end{proof}

The following proposition introduces an important feature of stable models.

\begin{proposition}\label{prop:mfp}
	Any stable model of a MANLP $\P$ is a minimal fix-point of $T_\P$.
\end{proposition}
\begin{proof}
	We will prove that the immediate consequence operator of a MANLP $\P$ coincides with the immediate consequence operator of the positive multi-adjoint logic program $\P_I$, for any $L$-interpretation $I$.
	
	Given a rule $\langle p\leftarrow_i @[ p_1, \dots, p_m,\neg p_{m+1},\dots,\neg p_n]; \vartheta \rangle$ in $\P$, for each $L$-interpretation $I$, we obtain the following chain of equalities:
{\footnotesize	
\begin{eqnarray*}
	\vartheta\Dotted{\adjoint_i} \hat{I}({@[ p_1, \dots, p_m,\neg p_{m+1},\dots,\neg p_n]})\}
	\!\!\!\!&=&\!\!\!\!\vartheta\Dotted{\adjoint_i} {\Dotted @[ {I}(p_1), \dots,{I} (p_m), \Dotted\neg{I}( p_{m+1}),\dots, \Dotted \neg{I}(p_n)}]\\
	\!\!\!\!&=&\!\!\!\!\vartheta\Dotted{\adjoint_i} {\Dotted@_{I}[ {I}(p_1), \dots, {I} (p_m)]}\\
	\!\!\!\!&=&\!\!\!\!\vartheta\Dotted{\adjoint_i}\hat{I}( @_{I}[ p_1, \dots, p_m])
\end{eqnarray*}	}
where $\langle p\leftarrow_i @_{I}[ p_1, \dots, p_m]; \vartheta \rangle$ is a rule in $\P_I$. Applying the supremum in both sides of the previous equality, we have that:
		\begin{eqnarray*}
	T_\P(I)(p)&=&\sup\{\vartheta\Dotted{\adjoint_i} \hat{I}(\mathcal{B}) \mid \langle p\leftarrow_i \mathcal{B};\vartheta \rangle\in\P \}\\
	&=&\sup\{\vartheta\Dotted{\adjoint_i} \hat{I}(\mathcal{B_I}) \mid \langle p\leftarrow_i \mathcal{B_I};\vartheta \rangle\in\P_I \} =T_{\P_I}(I)(p)
	\end{eqnarray*}
	\noindent
	for all $L$-interpretation $I$.
	
	Now, we will consider a stable model $M$ of $\P$, which is a minimal model of the positive multi-adjoint program $\P_M$, by Definition~\ref{def:stable}. Taking into account Proposition~\ref{monot:ico} and Knaster-Tarski's fix-point theorem, we can assert that $M$ is a fix-point  of $T_{\P_M}$. As the equality $T_\P=T_{\P_M}$ holds, we can conclude that $M=T_{\P_M}(M)=T_\P(M)$ and therefore $M$ is a fix-point of $T_\P$.
	
	It only remains to demonstrate the minimality of $M$. Let us assume a fix-point $N$ of $T_\P$ satisfying that $N\preceq M$. Then, by Proposition~\ref{prop6}, we obtain that $N$ is a model of $\P$. Moreover, by Proposition~\ref{prop:2.11NicoyManolo}, we have that each stable model of $\P$ is a minimal model of $\P$. Therefore, we conclude that $N=M$.
	\qed
\end{proof}

In general, the counterpart of Proposition~\ref{prop:mfp} is not true because the $T_\P$ operator is not necessarily monotonic.
	
\section{On the existence and unicity of stable models}\label{sec:exi-uni}

Interesting results about the existence and unicity of stable models for normal residuated logic programs were presented in \cite{escim:stablemodels} and~\cite{Madrid}. The aim of this section is to study which conditions are required in order to:

\begin{enumerate}
\item  generalize the existence of stable models for MANLPs defined on any convex compact set of an euclidean space; and
\item ensure the uniqueness of a stable model for a MANLP defined on the set of subintervals of   $[0,1]\times[0,1]$, which is denoted as $\mathcal{C}([0,1])=\{[x,y]\in[0,1]\times[0,1]\mid x\leq y\}$, together with the ordering relation $\leq$ defined as $[a,b]\leq[c,d]$ if and only if $a\leq c$ and $b\leq d$, for all $[a,b],[c,d]\in \mathcal{C}([0,1])$.
\end{enumerate}

\subsection{Existence of stable models in convex compact sets}\label{sec:exstable}
	
First of all, we will prove some properties of the set of interpretations. Given a finite multi-adjoint normal logic program $\P$ defined on $K$, the set of interpretations $\cal{I}_\mathfrak{K}$, together with the ordering relation defined in Section~\ref{sec:semlanguage}, verifies some properties of the underlying lattice. For example, by Proposition~\ref{prop:set:inter}, we can ensure that $(\cal{I}_\mathfrak{K},\sqsubseteq)$  is a complete lattice. Note that, each $K$-interpretation can be seen as an element of $K^n$, where $n$ is the cardinal of $\Pi_\P$. As a consequence, the set of $K$-interpretations inherits the properties of $K$ by means of the cartesian product.

In what follows, we will demonstrate that the whole set of interpretations of a MANLP defined on a lattice with convex (closed, respectively) carrier is a convex ({compact}, respectively) set.

\begin{proposition}\label{prop:convexandclosed}
Let $\P$ be a MANLP defined on a multi-adjoint normal lattice $(K,\preceq,\leftarrow_1,\adjoint_1,\dots,\leftarrow_n,\adjoint_n,\neg )$ where $K$ is a convex (closed, resp.) set of an euclidean space $X$. Then the set of $K$-interpretations of $\P$ is a convex  ({compact, resp.}) set in the set of mappings  defined on $X$.
\end{proposition}
	
\begin{proof}
Given the euclidean space of functions from $\Pi_\P$ to $K$ with the ordering relation $\sqsubseteq$ defined on the set of $K$-interpretations, and two $K$-interpretations $I,J\in I_\mathfrak{K}$, we will demonstrate that $t I+(1-t)J\in I_\mathfrak{K}$, for all $t\in[0,1]$. Since $I(p),J(p)\in {K}$, for all $p\in\Pi_\P$, we obtain that $t I(p)+(1-t)J(p)\in K$, and therefore $t I+(1-t)J\in I_\mathfrak{K}$, for all $t\in[0,1]$. As a consequence, we conclude that $\mathit{I}_\mathfrak{K}$ is a convex set.
		
In order to prove that $\mathit{I}_\mathfrak{K}$ is a compact set, we need to demonstrate that the set of $K$-interpretations of $\P$ is bounded and closed.  Since $(K,\preceq)$ is a bounded lattice, with   bottom  and   top  elements $\bot$ and $\top$, respectively, we can define the constant  bottom and top interpretation   $I_\bot$ and $I_\top$. Taking into account the ordering relation defined on the set of interpretations, we obtain that $I_\bot \sqsubseteq  I \sqsubseteq I_\top$, for all $K$-interpretation $I$. Therefore, $\mathit{I}_\mathfrak{K}$ is a bounded set.
{On the other hand, we can ensure that $\mathit{I}_\mathfrak{K}$ is also closed since each $K$-interpretation can be seen as an element of $K^n$, where $n$ is the cardinal of $\Pi_\P$, and the cartesian product of closed sets is closed.}
 \qed
\end{proof}

Notice that,  if $K$ is closed then it is a compact set, since we are considering a multi-adjoint normal lattice. Hence, from now on, in order to not create confusion we will write compact instead of closed.

Now, we will show the considered mathematical reasoning to demonstrate the theorem related to the existence of stable models in MANLPs. Our purpose is to prove the continuity of the operator $R$ defined by $R(I)=\text{lfp}(T_{\P_I})$, where $\text{lfp}(T_{\P_I})$ is the least fix-point of the operator $T_{\P_I}$, for a given fuzzy $K$-interpretation $I$ where $K$ is a convex compact set. This fact allows us to apply Theorem \ref{Schauder} and therefore, we can guarantee that  an interpretation $I$ exists such that it is the least fix-point of $T_{\P_I}$. Furthermore, this fix-point $I$ is the least model of the positive multi-adjoint logic program $\P_I$. Therefore, we can ensure that $I$ is a stable model of~$\P$.

In order to reach this purpose, it will be fundamental to require the continuity of the conjunction connectives, the negation operator and the aggregator operators appearing in the body of the rules of MANLPs.

\begin{theorem}\label{thm:existencia}
Let $(K,\preceq,\leftarrow_1,\adjoint_1,\dots,\leftarrow_n,\adjoint_n,\neg )$ be a multi-adjoint normal lattice where $K$ is a non-empty convex compact set in an euclidean space and $\P$ be a finite MANLP defined on this lattice. If $\adjoint_1,\dots, \adjoint_n$, $\neg$ and the aggregator operators in the body of the rules of  $\P$ are continuous operators, then $\P$  has at least a stable model.
\end{theorem}	
\begin{proof}
Given a MANLP $\P$ and a $K$-interpretation $I$, the operator $R(I)=\text{lfp}(T_{\P_I})$ can be expressed as a composition of the operators $\mathcal{F}_1(I)=\P_I$ and $\mathcal{F}_2(\P)=\text{lfp}(T_{\P})$.

Note that $\mathcal{F}_1$ is a mapping from the set of $K$-interpretations to $(\mathfrak{R}^+_{\Pi\uplus K})^k=\mathfrak{R}^+_{\Pi\uplus K}\times\stackrel{k)}\dots\times\mathfrak{R}^+_{\Pi\uplus K}$, where $k$ is the number of rules in $\P$. For each $K$-interpretation $I$, we obtain:
$$
\mathcal{F}_1(I)\!=\!(p^1\!\!\leftarrow_{j_1}\!\!\!@_I^1[p_1^1,\dots, p_m^1],\dots,p^k\!\!\leftarrow_{j_k}\!\!\!@_I^k[p_1^k,\dots, p_m^k])
$$
where $j_i\in\{1,\dots, n\}$, with $i\in\{1,\dots,k\}$. Hence, $\mathcal{F}_1$ is a continuous mapping if and only if each component of $\mathcal{F}_1$ is continuous. But this is trivial since $@^i_I$ are continuous operators, by hypothesis. 

On the other hand, $\mathcal{F}_2$ is a mapping from $(\mathfrak{R}^+_{\Pi\uplus K})^k$ to the set of $K$-interpretations. Since every operator used in the computation of $T_\P$ is continuous, we can ensure that the immediate consequence operator is continuous. In addition, taking into account Proposition 5.4 in~\cite{Ll}, we can obtain the least fix-point of $T_\P$ by iterating $\omega$ times the inmediate consequence operator from the bottom interpretation. Hence, $\mathcal{F}_2$ is a continuos operator since it is a numerable composition of continuous operators.
		
Consequently, $R(I)=\text{lfp}(T_{\P_I})$ is continuous because it is  composition of two continuous operators. Applying Theorem \ref{Schauder} to the operator $R$, we conclude that $R$ has a  fix-point. Moreover, this fix-point coincides with the least model of $\P_I$ since it is a positive multi-adjoint logic program and we can apply Proposition~\ref{prop6}. Thus, it is a stable model of $\P$.\qed
\end{proof}

The following examples illustrate the result obtained in Theorem~\ref{thm:existencia}.

\begin{example}\label{ex_uni}
Consider the euclidean space $(X,\oplus,\otimes,\mathbb{R})$ where $X$ is the space of triangular functions defined by:
$$
f_n(z)= \left\{ \begin{array}{llc}
10(z-n)+1 &  \quad\hbox{if} & \hspace{-1cm}n-0.1\leq z\leq n \\
10(n-z)+1 &  \quad\hbox{if} & \hspace{-1cm}n\leq z < n+0.1 \\
0 &  \quad\hbox{otherwise} &
\end{array}
\right.
$$
with $n\in\mathbb{R}$. The operations $\oplus, \otimes\colon X \rightarrow X$ are defined as $f_n\oplus f_m=f_{n+m}$ and $k\otimes f_n=f_{k\cdot n}$, respectively, where $n,m,k\in \mathbb{R}$.

Now, we will consider the set of functions $K=\{f_x\mid x\in[0,1] \}$ together with the following ordering relation: $f_n\leq f_m$ if and only if $n\leq m$, for all $n,m\in\mathbb{R}$.

In order to see that $K$ is a convex set, for all $f_x,f_y\in K$ and $t\in [0,1]$, we will prove that $t \otimes f_x\oplus(1-t)\otimes f_y\in K$, which is equivalent to demonstrate that $f_{t\cdot x+(1-t)\cdot y}\in K$. Clearly, $t\cdot x+(1-t)\cdot y\in[0,1]$ since $x,y,t \in[0,1]$. Therefore, we obtain that $f_{t\cdot x+(1-t)\cdot y}\in K$ and consequently $K$ is convex.

From the ordering relation defined previously, we can assert that $K$ is a bounded set because $f_x\leq f_1$, for all $f_x\in K$. Furthermore, $K$ is a closed set since the boundary of $K$ is contained by it, that is, $\{f_0,f_1\}\subseteq K$.

Hence, we can conclude that $K$ is a convex compact set in $X$. Therefore, Theorem~\ref{thm:existencia} ensures that every multi-adjoint normal logic program $\P$ defined on the multi-adjoint normal lattice $(K,\leq,\leftarrow_1,\adjoint_1,\dots,\leftarrow_n,\adjoint_n,\neg )$, where the conjunctors, the negation and the aggregator operators in the body of the rules of  $\P$ are continuous, has at least a stable model.
\qed
\end{example}

Analogously, we can consider the functions $f_n$ with any width different from $0.1$. This kind of functions are interpreted as fuzzy numbers. A similar example can be obtained when we consider the set of functions $f_x$ where $x$ is an element of an arbitrary convex compact set.
	
The following example considers  another algebraic structure with  a more general family of triangular functions.

\begin{example}
Let $(X,\oplus,\otimes,\mathbb{R})$ be an euclidean space such that $X$ is composed of the triangular functions $f_{a_1,a_2,a_3}$ defined as follows:
$$
f_{a_1,a_2,a_3}(z)= \left\{ \begin{array}{llc}
\frac{z-a_1}{a_2-a_1} &  \quad\hbox{if} & \hspace{-1cm}a_1\leq z\leq a_2 \\
\frac{a_3-z}{a_3-a_2} &   \quad\hbox{if}  & \hspace{-1cm}a_2\leq z\leq a_3  \\
0 &  \quad\hbox{otherwise} &
\end{array}
\right.
$$
where $a_1,a_2,a_3\in\mathbb{R}$. For all $a_i,b_i,k\in \mathbb{R}$ with $i\in\{1,2,3\}$,
we define the operations $\oplus, \otimes\colon X \rightarrow X$ as:
\begin{eqnarray*}
f_{a_1,a_2,a_3}\oplus f_{b_1,b_2,b_3}&=&f_{a_1+b_1,a_2+b_2,a_3+b_3}\\
k\otimes f_{a_1,a_2,a_3}&=&f_{k\cdot a_1,k\cdot a_2,k\cdot a_3}
\end{eqnarray*}
which clearly are well defined. Following an analogous reasoning to the previous example and considering the ordering relation $f_{x_1,x_2,x_3}\leq f_{y_1,y_2,y_3}$ if and only if $x_1\leq y_1$, $x_2\leq y_2$ and $x_3\leq y_3$, we can ensure that $K=\{f_{x_1,x_2,x_3}\mid x_1,x_2,x_3\in[0,1] \}$ is a convex compact set. Once again, we can assert that every multi-adjoint normal logic program defined on $(K,\leq,\leftarrow_1,\adjoint_1,\dots,\leftarrow_n,\adjoint_n,\neg )$, such that $\adjoint_1,\dots,\adjoint_n, \neg$ and the aggregator operators in the body of the rules of  $\P$ are continuous operators, has at least a stable model.\qed
\end{example}

As usual, the existence theorem does not ensure the uniqueness of stable models, as we will show next.
\begin{example}
We will consider the following MANLP $\P$, defined on the multi-adjoint normal lattice $\langle[0,1],\leq,\leftarrow_G,\adjoint_G,\leftarrow_P,\adjoint_P,\neg\rangle$, with five rules and one fact.
\begin{equation*}
\begin{array}{ll}
r_1:\ \langle p\leftarrow_G \neg t\ \ ;\ 0.6\rangle &
r_4:\ \langle t\leftarrow_P s\ ;\ 1\rangle\\
r_2:\ \langle q\leftarrow_P \neg s\ ;\ 0.8\rangle &
r_5:\ \langle s\leftarrow_P 1\ ;\ 0.5\rangle\\
r_3:\ \langle p\leftarrow_P q\ \adjoint_P s\ ;\ 0.9\rangle &
r_6:\ \langle t\leftarrow_G \neg q\ \adjoint_G\neg p\ ;\ 0.7\rangle
\end{array}
\end{equation*}
Notice that, the set of propositional symbols is given by $\Pi_\P=\{p,q,s,t\}$ and the operators included in the multi-adjoint normal lattice are the Gödel and product conjunctors, $\adjoint_G$ and $\adjoint_P$, together with their corresponding adjoint implications, $\leftarrow_G$ and $\leftarrow_P$. The considered negation operator is the standard negation defined as $\neg(x)=1-x$, for each $x\in[0,1]$.

Clearly, $[0,1]$ is a convex compact set and the operators in the body of the rules of  $\P$ are continuous. Therefore, applying Theorem~\ref{thm:existencia}, we can guarantee that the multi-adjoint normal logic program $\P$ defined on the multi-adjoint normal lattice $\langle[0,1],\leq,\leftarrow_G,\adjoint_G,\leftarrow_P,\adjoint_P,\neg\rangle$ has at least a stable model. In the following, we will compute two different stable models.

From the interpretation $M\equiv\{(p,0.4),(q,0.4),(s,0.5),(t,0.6)\}$, we can define the corresponding reduct $\P_M$ as follows:
\begin{equation*}
\begin{array}{ll}
r^M_1:\ \langle p\leftarrow_G 0.4 \ ;\ 0.6\rangle &
r^M_4:\ \langle t\leftarrow_P s\ ;\ 1\rangle\\
r^M_2:\ \langle q\leftarrow_P 0.5 \ ;\ 0.8\rangle &
r^M_5:\ \langle s\leftarrow_P 1\ ;\ 0.5\rangle\\
r^M_3:\ \langle p\leftarrow_P q\ \adjoint_P s\ ;\ 0.9\rangle &
r^M_6:\ \langle t\leftarrow_G 0.6\ ;\ 0.7\rangle
\end{array}
\end{equation*}

First of all, we compute the   least model of the program $\P_M$. For that, since  $\P_M$ is a positive program, we iterate the $T_{\P_M}$ operator from the minimum interpretation $I_\bot$.
\begin{center}
\begin{tabular}{|c||c|c|c|c|}
\hline
  & $p$ & $q$ & $s$ & $t$\\[1ex]
\hline
$I_\bot$ & 0 &0 & 0 & 0\\
\hline
$T_{\P_M}(I_\bot)$ & 0.4 & 0.4 & 0.5 & 0.6\\[0.5ex]
\hline
$T_{\P_M}^2(I_\bot)$ & 0.4 & 0.4 & 0.5 & 0.6\\[0.5ex]
\hline
\end{tabular}
\end{center}

Consequently, since $T_{\P_M}(I_\bot)=M$ and it is the least fix-point of $T_{\P_M}$, $M$ is the least model of the reduct $\P_M$, which allows us to ensure that $M$ is a stable model of the program $\P$.

Now, we will show that $M$ is not the unique stable model of $\P$. Let $M'$ be the interpretation given by
$M'\equiv\{(p,0.5),(q,0.4),(s,0.5),(t,0.5)\}$
Then, the corresponding reduct $\P_{M'}$ is defined as:
\begin{equation*}
\begin{array}{ll}
r^{M'}_1:\ \langle p\leftarrow_G 0.5 \ ;\ 0.6\rangle &
r^{M'}_4:\ \langle t\leftarrow_P s\ ;\ 1\rangle\\
r^{M'}_2:\ \langle q\leftarrow_P 0.5\ ;\ 0.8\rangle &
r^{M'}_5:\ \langle s\leftarrow_P 1\ ;\ 0.5\rangle\\
r^{M'}_3:\ \langle p\leftarrow_P q\ \adjoint_P s\ ;\ 0.9\rangle &
r^{M'}_6:\ \langle t\leftarrow_G 0.5\ ;\ 0.7\rangle
\end{array}
\end{equation*}
By an analogous reasoning to that given for the least model of the reduct $\P_M$, it can be easily proved that $M'$ is the least model of $\P_{M'}$. Therefore, $M'$ is also a stable model of the program $\P$.\qed
\end{example}

Hence, ensuring the  existence of a unique stable model is an important challenge. For example, when the logic program is associated with a search problem, if the stable model is unique, then  the problem is solvable and it has a unique solution, which determines the  optimal information we can   obtain  from the knowledge system. Therefore,  studying sufficient conditions in order to ensure the uniqueness is an important goal, which will be developed in the next section.

\subsection{Unicity of stable models in MANLPs defined on $\mathcal{C}([0,1])$}\label{sec:unicity}

As we argued above and in the introduction section, the characterization of programs with a unique stable model is important.
This section will consider a special algebraic structure and sufficient conditions from which we can ensure the unicity of stable models for  multi-adjoint normal logic programs defined on  the set of subintervals of $[0,1]\times[0,1]$, which is denoted by  $\mathcal{C}([0,1])$.

In the following, we will present the particular operators which are considered in the programs.

\begin{definition}\label{def:intervalconjunctor}
Given $\alpha,\beta,\gamma,\delta\in \mathbb{N}$ such that $\beta\leq \alpha$ and $\delta\leq \gamma$, the operator $\adjoint_{\beta\delta}^{\alpha\gamma}\colon \mathcal{C}([0,1])^2\rightarrow \mathcal{C}([0,1])$ defined as:
$$
\adjoint_{\beta\delta}^{\alpha\gamma}([a,b],[c,d])=[\,a^\alpha* c^\gamma,\, b^\beta*d^\delta\,]
$$
with $a,b,c,d\in \mathbb{R}$ and $*$ being the usual product among real numbers, will be called \emph{exponential interval product with respect to $\alpha, \beta,\gamma$ and $\delta$} (\emph{ei-product}, in short).
\end{definition}

Note that  every ei-product  with respect to four natural numbers $\alpha, \beta,\gamma$ and $\delta$ is well defined, since these values   satisfy that  $\beta\leq \alpha$ and $\delta\leq \gamma$.

In~\cite{ipmu10-medina}, different properties of these operators were introduced. In particular, the existence of the residuated implication $\leftarrow_{\beta\delta}^{\alpha\gamma}$ was proved (see~\cite[Theorem 1]{ipmu10-medina} for more details). Hence, we have that $(\adjoint_{\beta\delta}^{\alpha\gamma},\leftarrow_{\beta\delta}^{\alpha\gamma})$ forms an adjoint pair  and so, they can be used in any  multi-adjoint normal logic program.

An extension on $\mathcal{C}([0,1])$ of the standard negation will be the negation operator that we will consider in the programs. Specifically, this operator      $\neg\colon\mathcal{C}([0,1])\rightarrow\mathcal{C}([0,1])$  will be defined as $\neg[a,b]=[1-b,1-a]$ for all $[a,b]\in\mathcal{C}([0,1])$, which is clearly    decreasing and satisfies  $\neg[0,0]=[1,1]$ and  $\neg[1,1]=[0,0]$.

Before presenting the results associated with the unicity of stable models, we will take into consideration the following remarks. Due to the relation between  $\mathcal{C}([0,1])$ and $ [0,1]\times [0,1]$, we can introduce the inclusion mapping $\iota\colon \mathcal{C}([0,1])\to [0,1]\times [0,1]$, defined as $\iota([a,b])=(a,b)$.  This mapping can easily be extended to tuples as follows,  $\iota\colon \mathcal{C}([0,1])^n\to [0,1]^n\times [0,1]^n$, defined by $\iota([a_1,b_1],\dots, [a_n,b_n])=(a_1,\dots, a_n,b_1, \dots, b_n)$.

On the other hand, given the propositional symbols $p_1,\ldots,p_n$ appearing in $\P$, we can express each $\mathcal{C}([0,1])$-interpretation $I$ as a tuple $(I(p_1),\!\dots\!,I(p_n))$ which belongs to $(\mathcal{C}([0,1]))^n$. Therefore, the mapping $\iota$ can be defined on the set of  $\mathcal{C}([0,1])$-interpretations and the image of each $\mathcal{C}([0,1])$-interpretation will be a n-tuple, which will be denoted with a bar, that is, given a  $\mathcal{C}([0,1])$-interpretation $I$, we will write $\iota (I)=\bar I$.

For example, given $\Pi=\{p,q,s\}$ and the  $\mathcal{C}([0,1])$-interpretation $I\colon \Pi \to  \mathcal{C}([0,1])$, defined as $I(p)=[0.1,0.4]$,  $I(q)=[0,0]$, and $I(s)=[0.7,0.9]$, if we consider the alphabetical ordering   among the propositional symbols, $I$ can be written as the tuple $\bar I=([0.1,0.4], [0,0], [0.7,0.9])$.\label{comentariointerpret}

Moreover, $T_\P$ can be considered as a real function from $\mathcal{C}([0,1])^{n}$ to $\mathcal{C}([0,1])^{n}$
since  it assigns $\mathcal{C}([0,1])$-interpretations to $\mathcal{C}([0,1])$-interpretations. Hence, we will write $T_\P(I)(p_i)=(T_\P)_i(I)$ in order to express the value of $T_\P(I)$ for each propositional symbol $p_i$ and we also have $(T_\P)_i(I)=[(T_\P)^1_i(I),(T_\P)^2_i(I)]$ since $(T_\P)_i(I)\in\mathcal{C}([0,1])$.
 Considering the mapping $\iota$, we can write the n-tuple $([a_1,b_1],\dots, [a_n,b_n]) $ in $\mathcal{C}([0,1])^{n}$ as $(a_1,\dots, a_n,b_1, \dots, b_n)$. This fact allows us to define the mapping $\overline{T_\P}\colon\iota(\mathcal{I}_\mathfrak{L})\rightarrow [0,1]^n\times[0,1]^n$ as $\overline{T_\P}(\iota(I))=\iota(T_\P(I))$, for each $I\in\mathcal{I}_\mathfrak{L}$. Notice that $\iota(\mathcal{I}_\mathfrak{L})=\{\iota(I)\mid I\in \mathcal{C}([0,1])^{n}\}$.

Taking into account these previous considerations, we will introduce a lemma required to prove the uniqueness of the stable models for multi-adjoint normal programs defined  on $\mathcal{C}([0,1])$  by using the extension of the standard negation defined above and a family of adjoint pairs formed by different ei-products together with their corresponding residuated implications.

\begin{lemma}\label{lem:uni}
	Let $\P$ be a MANLP defined on a multi-adjoint normal lattice $(\mathcal{C}([0,1]),\leq,\leftarrow_{\beta_{1}\delta_{1}}^{\alpha_{1}\gamma_{1}}, \adjoint_{\beta_{1}\delta_{1}}^{\alpha_{1}\gamma_{1}},\dots,\leftarrow_{\beta_{m}\delta_{m}}^{\alpha_{m}\gamma_{m}}, \adjoint_{\beta_{m}\delta_{m}}^{\alpha_{m}\gamma_{m}}, \neg )$ such that at most one rule with head $p$ appears in $\P$ and the only possible operators in the body of the rules\footnote{Notice that the implications in the rules can be the residuated implications of any general ei-product.} are $\adjoint_{\beta\delta}^{\alpha\gamma}$ with $\alpha=\beta=\gamma=\delta=1$. If $I=[I^1, I^2]$ and $J=[J^1, J^2]$ are two $\mathcal{C}([0,1])$-interpretations, such that $J\sqsubseteq I$, then:
$$
\sum_{l=1}^{2}\sum_{j=1}^{n}\bigg|\frac{\partial(\overline{T_\P})^1_i}{\partial p_j^l}(J^1(p_1),\dots,J^1(p_n),J^2(p_1),\dots,J^2(p_n))\bigg|\leq \alpha
$$
$$
\sum_{l=1}^{2}\sum_{j=1}^{n}\bigg|\frac{\partial(\overline{T_\P})^2_i}{\partial p_j^l}(J^1(p_1),\dots,J^1(p_n),J^2(p_1),\dots,J^2(p_n))\bigg|\leq \beta
$$
where
\begin{eqnarray*}
\alpha&=&\sum_{j=1}^{h}(\vartheta^1)^{\alpha_w}\cdot \gamma_w\cdot I^1(q_j)^{\gamma_w-1}\cdot(I^1(q_1)\cdots I^1(q_{j-1})\cdot I^1(q_{j+1})\cdots I^1(q_h))^{\gamma_w}\\
& & +\ (\vartheta^1)^{\alpha_w}\cdot \gamma_w\cdot(k-h)(I^1(q_1)\cdots I^1(q_h))^{\gamma_w}\\
\beta&=&\sum_{j=1}^{h}(\vartheta^2)^{\beta_w}\cdot \delta_w\cdot I^2(q_j)^{\delta_w-1}\cdot(I^2(q_1)\cdots I^2(q_{j-1})\cdot I^2(q_{j+1})\cdots I^2(q_h))^{\delta_w}\\
& & +\ (\vartheta^2)^{\beta_w}\cdot \delta_w\cdot(k-h)(I^2(q_1)\cdots I^2(q_h))^{\delta_w}
\end{eqnarray*}

	\noindent
and $\langle p_i\leftarrow_{\beta_{w}\delta_{w}}^{\alpha_{w}\gamma_{w}} q_1*\cdots*q_h*\neg q_{h+1}*\cdots*\neg q_k;\ [\vartheta^1,  \vartheta^2]\rangle$, with $w\in\{1,\dots,m\}$, is the unique rule in $\P$ with head $p_i$.
\end{lemma}

\begin{proof}
Let us assume that only one rule in $\P$ with head $p_i$ exists, that is,
$\langle p_i\leftarrow_{\beta_{w}\delta_{w}}^{\alpha_{w}\gamma_{w}} q_1*\cdots*q_h*\neg q_{h+1}*\cdots*\neg q_k;\ [\vartheta^1,  \vartheta^2]\rangle$. Hence, we have that $(T_\P)_i(I)$ is equal to:
$$
 [\vartheta^1,  \vartheta^2]\Dotted{\adjoint}_{\beta_{w}\delta_{w}}^{\alpha_{w}\gamma_{w}}\hat{I}(q_1*\cdots*q_h*\neg q_{h+1}*\cdots*\neg q_k)
$$
Note that, by definition of the negation operator  used here, we obtain that:
{\footnotesize \begin{eqnarray*}
	\hat{I}^1(q_1\!*\cdots*\!q_h*\neg q_{h+1}\!*\cdots*\!\neg q_k)\!\!\!\!& = &\!\!\!\! I^1(q_1)\!*\cdots*\!I^1(q_h)\!*\!(1-I^2(q_{h+1}))\!*\cdots*\!(1-I^2(q_k))\\[10pt]
	\hat{I}^2(q_1\!*\cdots*\!q_h*\neg q_{h+1}\!*\cdots*\!\neg q_k)\!\!\!\!& = &\!\!\!\! I^2(q_1)\!*\cdots*\!I^2(q_h)\!*\!(1-I^1(q_{h+1}))\!*\cdots*\!(1-I^1(q_k))
\end{eqnarray*}}
Therefore, considering each component of the immediate consequence operator and Definition~\ref{def:intervalconjunctor}, we have that:
 \begin{eqnarray*}
(T_\P)^1_i(I) \!\!\!\!& = &\!\!\!\! (\vartheta^1)^{\alpha_w}*\!\big(I^1(q_1)\!*\cdots*\!I^1(q_h)\!*\!(1-I^2(q_{h+1}))\!*\cdots*\!(1-I^2(q_k))\big)^{\gamma_w}\\[10pt]
(T_\P)^2_i(I) \!\!\!\!& = &\!\!\!\! (\vartheta^2)^{\beta_w}*\!\big(I^2(q_1)\!*\cdots*\!I^2(q_h)\!*\!(1-I^1(q_{h+1}))\!*\cdots*\!(1-I^1(q_k))\big)^{\delta_w}
\end{eqnarray*}
Now, by using the mapping $\iota$,
the first component of the immediate consequence operator can be written as\footnote{Notice that the variables have been denoted with the propositional symbols, abusing of notation.}:
$$
(\overline{T_\P})^1_i(p_1^1,\dots,p_n^1,p_1^2,\dots,p_n^2)=(\vartheta^1)^{\alpha_w}\cdot\big(q_1*\cdots*q_k*(1-q_{k+1})*\cdots*(1-q_m)\big)^{\gamma_w}
$$
\noindent{where each $q_j$ with $j\leq k$ is actually some $p_1^1,\dots,p_n^1$ and each $q_j$ with $j> k$ is some $p_1^2,\dots,p_n^2$.}

Analogously, the second component of the immediate consequence operator can be expressed as:
$$
(\overline{T_\P})^2_i(p_1^1,\dots,p_n^1,p_1^2,\dots,p_n^2)=(\vartheta^2)^{\beta_w}\cdot\big(q_1*\cdots*q_k*(1-q_{k+1})*\cdots*(1-q_m)\big)^{\delta_w}
$$
\noindent{where each $q_j$ with $j\leq k$ is actually some $p_1^2,\dots,p_n^2$ and each $q_j$ with $j> k$ is some $p_1^1,\dots,p_n^1$.}

Since  $(\overline{T_\P})^1_i$ and $(\overline{T_\P})^2_i$ are composition of differentiable mappings, they are also differentiable mappings. Considering only $(\overline{T_\P})^1_i$, we will  compute its partial derivatives distinguishing different cases:
\begin{itemize}
\item [(a)] If $p_j=q_t$, with $t\leq h$, then
	{\small\begin{eqnarray*}
		 \frac{\partial(\overline{T_\P})^1_i}{\partial p_j}\!\!\!\!&=&\!\!\!\!(\vartheta^1)^{\alpha_w} \cdot \gamma_{w} \cdot q_t^{\gamma_w-1}\cdot(q_1\cdots q_{t-1}\cdot q_{t+1}\!\cdots q_h\!\cdot\!(1\! -\! q_{h+1})\! \cdots\! (1\! -\! q_k))^{\gamma_w}
	\end{eqnarray*}}

\item [(b)]	When $p_j=q_t$, with $t>h$, we have:
	\small{\begin{eqnarray*}
	\frac{\partial(\overline{T_\P})^1_i}{\partial p_j}\!\!\!\!&=&\!\!\!\!(\vartheta^1)^{\alpha_w}\cdot (-\gamma_w)\cdot (1-q_t)^{\gamma_w-1}(q_1\cdots q_{h}\cdot (1-q_{h+1})\cdots (1-q_{t-1})\cdot\\
	& &(1-q_{t+1})\cdots(1-q_k))^{\gamma_w}
	\end{eqnarray*}}
\item [(c)] Otherwise, $\frac{\partial(\overline{T_\P})^1_i}{\partial p_j}=0$.
\end{itemize}
Note that, by the definition of the MANLP $\P$, all propositional symbols appearing in the body of a rule are different.

From the computations above, we obtain that the sum of all partial derivatives evaluated in $(J^1(p_1),\dots,J^1(p_n),J^2(p_1),\dots,J^2(p_n))$ verifies the next inequality:	
{\footnotesize	
\begin{eqnarray*}
	\lefteqn{	\sum_{l=1}^{2}\sum_{j=1}^{n}\bigg|\frac{\partial(\overline{T_\P})^1_i}{\partial p_j^l}(J^1(p_1),\dots,J^1(p_n),J^2(p_1),\dots,J^2(p_n))\bigg|=}\\
		&=&\!\!\!\!\sum_{j=1}^{h}\!\big|\vartheta^{\alpha_w}\!\!\cdot \gamma_w\!\!\cdot \!J^1(q_j)^{\gamma_w-1}\!\!\cdot\!\Big(J^1(q_1)\!\cdots\! J^1(q_{j-1})\!\cdot\! J^1(q_{j+1})\cdots J^1(q_h)\!\cdot\!(1-J^2(q_{h+1}))\cdots (1-J^2(q_k))\Big)^{\gamma_w}\big|\\[10pt]
		& &\!\!\!+\!\!\sum_{j=h+1}^{k}\big|\vartheta^{\alpha_w}\cdot (-\gamma_w)\cdot (1-J^2(q_j)^{\gamma_w-1})\cdot\!\Big(J^1(q_1)\cdots J^1(q_h)\cdot(1-J^2(q_{h+1}))\cdots
		\\[10pt]& &
		\cdots(1-J^2(q_{j-1}))\cdot\ (1-J^2(q_{j+1}))\cdots(1-J^2(q_k))\Big)^{\gamma_w}\big|\\[10pt]				
		&\leq& \!\!\!\!\Bigg(\sum_{j=1}^{h}\vartheta^{\alpha_w}\cdot \gamma_w\cdot J^1(q_j)^{\gamma_w-1}\cdot\Big(J^1(q_1)\cdots J^1(q_{j-1})\cdot J^1(q_{j+1})\cdots J^1(q_h)\Big)^{\gamma_w}\Bigg)\\
		& & +\sum_{j=h+1}^{k}\!\!\vartheta^{\alpha_w}\cdot \gamma_w\cdot(1-J^2(q_j)^{\gamma_w-1})(J^1(q_1)\cdots J^1(q_h))^{\gamma_w}\\[10pt]	
&\leq& \!\!\!\!\Bigg(\sum_{j=1}^{h}\vartheta^{\alpha_w}\cdot \gamma_w\cdot J^1(q_j)^{\gamma_w-1}\cdot\Big(J^1(q_1)\cdots J^1(q_{j-1})\cdot J^1(q_{j+1})\cdots J^1(q_h)\Big)^{\gamma_w}\Bigg)\\
	& &+\ \vartheta^{\alpha_w}\cdot \gamma_w\cdot(k-h)\cdot(J^1(q_1)\cdots J^1(q_h))^{\gamma_w}\\[10pt]
		&\leq& \!\!\!\!\Bigg(\sum_{j=1}^{h}\vartheta^{\alpha_w}\cdot \gamma_w\cdot I^1(q_j)^{\gamma_w-1}\cdot\Big(I^1(q_1)\cdots I^1(q_{j-1})\cdot I^1(q_{j+1})\cdots I^1(q_h)\Big)^{\gamma_w}\Bigg)\\
		& & +\ \vartheta^{\alpha_w}\cdot \gamma_w\cdot(k-h)\cdot(I^1(q_1)\cdots I^1(q_h))^{\gamma_w}
	\end{eqnarray*}}
An analogous reasoning with respect to the second component of the immediate consequence operator leads us to conclude that:
$$
\sum_{l=1}^{2}\sum_{j=1}^{n}\bigg|\frac{\partial(\overline{T_\P})^2_i}{\partial p_j^l}(J^1(p_1),\dots,J^1(p_n),J^2(p_1),\dots,J^2(p_n))\bigg|\leq \beta
$$
where
\begin{eqnarray*}
\beta&=&\sum_{j=1}^{h}(\vartheta^2)^{\beta_w}\cdot \delta_w\cdot I^2(q_j)^{\delta_w-1}\cdot(I^2(q_1)\cdots I^2(q_{j-1})\cdot I^2(q_{j+1})\cdots I^2(q_h))^{\delta_w}\\
& &+\ (\vartheta^2)^{\beta_w}\cdot \delta_w\cdot(k-h)(I^2(q_1)\cdots I^2(q_h))^{\delta_w}
\end{eqnarray*}\qed	
\end{proof}

Based on the previous results, the following theorem is focused on  the uniqueness of   stable models. The proof will be based on  demonstrating that $T_\P$ has only one fix-point in $\mathcal{C}([0,1])^n$. Considering Proposition~\ref{prop:mfp} and Theorem~\ref{thm:existencia}, we can state that each stable model of $\P$ is a minimal fix-point of $T_\P$ and there exists at least one stable model of $\P$, respectively. These facts will lead us to conclude that the only fix-point of $T_\P$ is the unique stable model.

\begin{theorem}\label{th:uni}
	Let $\P$ be a finite MANLP defined on the multi-adjoint normal lattice $(\mathcal{C}([0,1]),\leq,\leftarrow_{\beta_{1}\delta_{1}}^{\alpha_{1}\gamma_{1}}, \adjoint_{\beta_{1}\delta_{1}}^{\alpha_{1}\gamma_{1}},\dots,\leftarrow_{\beta_{m}\delta_{m}}^{\alpha_{m}\gamma_{m}}, \adjoint_{\beta_{m}\delta_{m}}^{\alpha_{m}\gamma_{m}}, \neg )$ such that the only possible operators in the body of the rules are $\adjoint_{\beta\delta}^{\alpha\gamma}$ with $\alpha=\beta=\gamma=\delta=1$, and $[\vartheta^1_p, \vartheta^2_p]\!=\!\max\{[\vartheta^1, \vartheta^2]\!\mid\!\langle p\leftarrow_{\beta_{w}\delta_{w}}^{\alpha_{w}\gamma_{w}} \mathcal{B} ;[\vartheta^1, \vartheta^2]\rangle\in\P\}$. If the inequality
{\footnotesize
$$
\sum_{j=1}^{h}(\vartheta^2)^{\beta_w}\cdot \delta_w\cdot (\vartheta^2_{q_j})^{\delta_w-1}\cdot\Big(\vartheta^2_{q_1}\!\cdots \vartheta^2_{q_{j-1}}\! \cdot\, \vartheta^2_{q_{j+1}}\cdots\vartheta^2_{q_h}\Big)^{\delta_w}\!\!+(\vartheta^2)^{\beta_w}\cdot \delta_w\cdot (k-h)(\vartheta^2_{q_1}\!\cdots \vartheta^2_{q_h})^{\delta_w}\!< \!1
$$}	
\noindent{holds for every rule $\langle p\leftarrow_{\beta_{w}\delta_{w}}^{\alpha_{w}\gamma_{w}} q_1*\cdots*q_h*\neg q_{h+1}*\cdots*\neg q_k; [\vartheta^1,  \vartheta^2]\rangle\in\P$, with $w\in\{1,\dots,m\}$, then there exists a unique stable model of $\P$.}
\end{theorem}
\begin{proof}
Given the $\mathcal{C}([0,1])$-interpretation $I_\vartheta$, which assigns the value $[\vartheta^1_p, \vartheta^2_p]$ to each propositional symbol $p\in\Pi_\P$,  the natural number $n$ representing the number of propositional symbols in $\Pi_\P$ and the set $A=\{\iota(J)\mid J\in{\mathcal{C}([0,1])}^n \hbox{ and } J\sqsubseteq I_\vartheta\}$, we will begin proving that   $\overline{ T_\P}$ is a contractive mapping in $A$ with respect to the supremum norm $||\cdot||_\infty$. That is, we will demonstrate that there exists a real value $0<\lambda<1$ such that:
\begin{equation}\label{def:contractiva}
||\overline{T_\P}(\overline{J_1})-\overline{T_\P}(\overline{J_2})||_\infty\leq||\overline{J_1}-\overline{J_2}||_\infty\cdot\lambda
\end{equation}
for each pair of  $\overline{J_1},\overline{J_2}\in A$. This fact will allow us to apply Banach fix-point theorem and to ensure that $\overline{T_\P}$ has only one fix-point in $A$. To reach this purpose, we distinguish two cases:

\textit{Base Case:} Given a multi-adjoint normal logic program $\P$, we will suppose that there exists at most one rule in $\P$ with head $p$, for each propositional symbol $p\in\Pi_\P$. In order to prove Equation~(\ref{def:contractiva}), we will apply the mean value theorem~\cite{TsoyWo} on each component of  $\overline{T_\P}=[\overline{T_\P}^1,\overline{T_\P}^2]$. First of all, we have to prove that the conditions of this theorem are satisfied. Considering $\overline{T_\P}^1$ and $\overline{T_\P}^2$ as functions defined on $\mathbb{R}^n$ in the way mentioned in Lemma~\ref{lem:uni}, both of them are differentiable functions in $\mathbb{R}^n$. Moreover, for all $\overline{J_1},\overline{J_2}\in A$, the line segment $S(\overline{J_1},\overline{J_2})=\{(1-t)\cdot \overline{J_1}+ t \cdot \overline{J_2}\mid 0\leq t \leq 1\}$ is contained in $A$. Hence, we can apply the mean value theorem on each component of  $\overline{T_\P}$, obtaining that:
\begin{equation}\label{tmavalormedio}
||\overline{T_\P}^1(\overline{J_1})-\overline{T_\P}^1(\overline{J_2})||_\infty\leq||\overline{J_1}-\overline{J_2}||_\infty\cdot\sup\{||D\overline{T_\P}^1(\overline{J})||_\infty\mid \overline{J}\in S(\overline{J_1},\overline{J_2})\}
\end{equation}
\begin{equation}\label{tmavalormedio2}
||\overline{T_\P}^2(\overline{J_1})-\overline{T_\P}^2(\overline{J_2})||_\infty\leq||\overline{J_1}-\overline{J_2}||_\infty\cdot\sup\{||D\overline{T_\P}^2(\overline{J})||_\infty\mid \overline{J}\in S(\overline{J_1},\overline{J_2})\}
\end{equation}
where
\begin{eqnarray*}
||D\overline{T_\P}^1(\overline{J})||_\infty&=&\sup\{||D\overline{T_\P}^1(\overline{J})(x)||_\infty\mid ||x||_\infty\leq 1\}\\
||D\overline{T_\P}^2(\overline{J})||_\infty&=&\sup\{||D\overline{T_\P}^2(\overline{J})(x)||_\infty\mid ||x||_\infty\leq 1\}\\
||x||_\infty&=&\max\{|x_i| \mid x=(x_1,\dots, x_n)\in \mathbb{R}^n\}
\end{eqnarray*}

In order to prove that $\overline{T_\P}^1$ is a contractive mapping in $A$, some previous considerations must be taken into account:	
\begin{enumerate}
\item[(a)] The line segment  $S(\overline{J_1},\overline{J_2})=\{(1-t)\cdot \overline{J_1}+ t \cdot \overline{J_2}\mid 0\leq t \leq 1\}$ is a compact set and consequently  $\sup\{||D\overline{T_\P}^1(\overline{J})||_\infty\mid \overline{J}\in S(\overline{J_1},\overline{J_2})\}$ is a maximum.

\item[(b)] Since only one rule with head $p$ exists in $\P$, the conditions required in Lemma~\ref{lem:uni} are satisfied and, therefore, for each $\overline{J}\in A$, the following inequalities hold:
{\small
\begin{eqnarray*}
\lefteqn{\sum_{l=1}^{2}\sum_{j=1}^{n}\bigg|\frac{\partial(\overline{T_\P})^1_i}{\partial p_j^l}(J^1(p_1),\dots,J^1(p_n),J^2(p_1),\dots,J^2(p_n))\bigg|\leq }\\[10pt]
&\leq& \sum_{j=1}^{h}(\vartheta^1)^{\alpha_w}\cdot \gamma_w\cdot I_\vartheta^1(q_j)^{\gamma_w-1}\cdot(I_\vartheta^1(q_1)\cdots I_\vartheta^1(q_{j-1})\cdot I_\vartheta^1(q_{j+1})\cdots I_\vartheta^1(q_h))^{\gamma_w} \nonumber \\
& & +\ (\vartheta^1)^{\alpha_w}\cdot \gamma_w\cdot(k-h)(I_\vartheta^1(q_1)\cdots I_\vartheta^1(q_h))^{\gamma_w} \nonumber \\
&=&\sum_{j=1}^{h}(\vartheta^1)^{\alpha_w}\cdot \gamma_w\cdot (\vartheta^1_{q_j})^{\gamma_w-1}\cdot\Big(\vartheta^1_{q_1}\cdots\vartheta^1_{q_{j-1}} \cdot\, \vartheta^1_{q_{j+1}}\cdots\vartheta^1_{q_h}\Big)^{\gamma_w} \nonumber \\
& & +\ (\vartheta^1)^{\alpha_w}\cdot \gamma_w\cdot (k-h)(\vartheta^1_{q_1}\cdots \vartheta^1_{q_h})^{\gamma_w} = \lambda^1 \nonumber
\end{eqnarray*}
}
An analogous reasoning can be given for $\overline{T_\P}^2$, obtaining:
{\small
\begin{eqnarray*}
\lefteqn{\sum_{l=1}^{2}\sum_{j=1}^{n}\bigg|\frac{\partial(\overline{T_\P})^2_i}{\partial p_j^l}(J^1(p_1),\dots,J^1(p_n),J^2(p_1),\dots,J^2(p_n))\bigg|\leq }\\[10pt]
&\leq& \sum_{j=1}^{h}(\vartheta^2)^{\beta_w}\cdot \delta_w\cdot I_\vartheta^2(q_j)^{\delta_w-1}\cdot(I_\vartheta^2(q_1)\cdots I_\vartheta^2(q_{j-1})\cdot I_\vartheta^2(q_{j+1})\cdots I_\vartheta^2(q_h))^{\delta_w} \\
& & +\ (\vartheta^2)^{\beta_w}\cdot \delta_w\cdot(k-h)(I_\vartheta^2(q_1)\cdots I_\vartheta^2(q_h))^{\delta_w} \\
&=&\sum_{j=1}^{h}(\vartheta^2)^{\beta_w}\cdot \delta_w\cdot (\vartheta^2_{q_j})^{\delta_w-1}\cdot\Big(\vartheta^2_{q_1}\cdots \vartheta^2_{q_{j-1}} \cdot\, \vartheta^2_{q_{j+1}}\cdots\vartheta^2_{q_h}\Big)^{\delta_w} +\\
& & +\ (\vartheta^2)^{\beta_w}\cdot \delta_w\cdot (k-h)(\vartheta^2_{q_1}\cdots \vartheta^2_{q_h})^{\delta_w} = \lambda^2
\end{eqnarray*}
}

\noindent
Taking into account the hypothesis, for each $\overline{J}\in A$, we have that $\lambda^2 < 1$ and so,
{\small
\begin{equation}\label{eq:segundacomponente}
\sum_{l=1}^{2}\sum_{j=1}^{n}\bigg|\frac{\partial(\overline{T_\P})^2_i}{\partial p_j^l}(J^1(p_1),\dots,J^1(p_n),J^2(p_1),\dots,J^2(p_n))\bigg|\leq \lambda^2 < 1
\end{equation}
}
According to the ordering defined on $\mathcal{C}([0,1])$ and considering that $\beta_w\leq \alpha_w$ y $\delta_w\leq \gamma_w$, we can ensure that $\lambda^1\leq\lambda^2$. Hence, by hypothesis, we deduce that:
{\small
\begin{equation}\label{eq:primeracomponente}
\sum_{l=1}^{2}\sum_{j=1}^{n}\bigg|\frac{\partial(\overline{T_\P})^1_i}{\partial p_j^l}(J^1(p_1),\dots,J^1(p_n),J^2(p_1),\dots,J^2(p_n))\bigg|\leq\lambda^1<1
\end{equation}
}

From Equation~(\ref{eq:primeracomponente}), for each $\overline{J}\in A$ and  $x\in \mathbb{R}^n$ such that $||x||_\infty\leq 1$, we obtain:
{\footnotesize
\begin{eqnarray*}
\lefteqn{||(D\overline{T_\P}^1(\overline{J}))(x)||_\infty=}\\[20pt]
&\stackrel{(1)}{=}&\!\!\!\!\max\limits\!\Bigg\{\sum_{l=1}^{2}\sum_{j=1}^{n}\bigg|\frac{\partial(\overline{T_\P})^1_i}{\partial p_j^l}(J^1(p_1),\dots,J^1(p_n),J^2(p_1),\dots,J^2(p_n))\!\cdot\! x_j\bigg|\mid i\in\{1,\dots,n\}\Bigg\}\\[15pt]
&\leq&\!\!\!\!\max\limits\!\Bigg\{\!\!\sum_{l=1}^{2}\sum_{j=1}^{n}\bigg|\frac{\partial(\overline{T_\P})^1_i}{\partial p_j^l}(J^1(p_1),\dots,J^1(p_n),J^2(p_1),\dots,J^2(p_n))\bigg|\mid i\in\{1,\dots,n\}\Bigg\}\\
&\leq&\lambda^1
\end{eqnarray*}}
for all $\overline{J}\in S(\overline{J_1},\overline{J_2})$, where $(1)$ is given by definition of $||\cdot||_\infty$ on matrices.
\end{enumerate}

Considering Equation~(\ref{tmavalormedio}), we can conclude that:
$$
\begin{array}{rcl}
||\overline{T_\P}^1(\overline{J_1})-\overline{T_\P}^1(\overline{J_2})||_\infty &\leq&||\overline{J_1}-\overline{J_2}||_\infty\cdot\sup\{||D\overline{T_\P}^1(\overline{J})||_\infty\mid \overline{J}\in S(\overline{J_1},\overline{J_2})\}\\[5pt]
&\stackrel{(a)}{\leq}&||\overline{J_1}-\overline{J_2}||_\infty\cdot\max\{||D\overline{T_\P}^1(\overline{J})||_\infty\mid  \overline{J}\in S(\overline{J_1},\overline{J_2})\}\\[5pt]
&\stackrel{(b)}{\leq}&||\overline{J_1}-\overline{J_2}||_\infty\cdot\lambda^1
\end{array}
$$
Therefore, $\overline{T_\P}^1$ is a contractive mapping in $A$ whose Lipschitz constant is $\lambda^1<1$. Following an analagous reasoning, we deduce that $\overline{T_\P}^2$ is a contractive mapping in $A$ with Lipschitz constant $\lambda^2<1$, that is:
$$
||\overline{T_\P}^2(\overline{J_1})-\overline{T_\P}^2(\overline{J_2})||_\infty\leq ||\overline{J_1}-\overline{J_2}||_\infty\cdot\lambda^2
$$

After proving  the contractivity of each one of the components of $\overline{T_\P}$ with respect to $||\cdot||_{\infty}$, we will prove that $\overline{T_\P}$ is contractive in $A$ with respect to $||\cdot||_{\infty}$. First of all, by definition of the norm $||\cdot||_{\infty}$, we obtain:

{\small
	\begin{eqnarray}\label{norma_por_componentes}
	||\overline{T_\P}(\overline{J_1})-\overline{T_\P}(\overline{J_2})||_\infty \!\!\!\!\!&=&\!\!\!\!\!
	\max\!\Big\{||\overline{T_\P}^1(\overline{J_1})-\overline{T_\P}^1(\overline{J_2})||_\infty,||\overline{T_\P}^2(\overline{J_1})-\overline{T_\P}^2(\overline{J_2})||_\infty\!\Big\}
	\end{eqnarray}
}
Therefore, by the contractivity of $\overline{T_\P}^1$ and $\overline{T_\P}^2$, the following chain holds:
$$
\begin{array}{rcl}
||\overline{T_\P}(\overline{J_1})-\overline{T_\P}(\overline{J_2})||_\infty  &\leq&\max\Big\{||\overline{J_1}-\overline{J_2}||_\infty\cdot\lambda^1,||\overline{J_1}-\overline{J_2}||_\infty\cdot\lambda^2\Big\}\\[5pt]
&\leq&\max\Big\{||\overline{J_1}-\overline{J_2}||_\infty,||\overline{J_1}-\overline{J_2}||_\infty\Big\}\cdot\max\{\lambda^1,\lambda^2\}\\[5pt]
&=& ||\overline{J_1}-\overline{J_2}||_\infty\cdot\max\{\lambda^1,\lambda^2\}
\end{array}
$$
Thus, the operator $\overline{T_\P}$ is a contractive mapping with Lipschitz constant $\max\{\lambda^1,\lambda^2\}=\lambda^2$, since $\lambda^1\leq\lambda^2$. Finally, by Banach fix-point theorem, we conclude that $\overline{T_\P}$ has only one fix-point in $A$.

\textit{General Case:} Now, we will prove the general case assuming that $\P$ is a general MANLP. By Proposition~\ref{prop:2.8NicoyManolo}, we can ensure the existence of a partition of MANLPs $\{\P_\gamma\}_{\gamma\in \Gamma}$ such that:

\begin{itemize}
\item Two rules with the same head for each $\P_\gamma$ do not exist.
\item It is satisfied that $T_\P(I)(p)=\sup\{T_{\P_\gamma}(I)(p)\mid \gamma\in \Gamma\}$.
\end{itemize}

Given a propositional symbol $p\in\Pi_\P$, we will denote by $\P_{\gamma^p_i}$, with $i\in\{1,\dots,m_p\}$, the subprograms of the partition with a unique rule whose head is $p$, where $m_p$ is the number of rules with head $p$.

Since $\P$ is a finite program and taking into account the properties of the partition, we obtain, for each $\mathcal{C}([0,1])$-interpretation $J$ and each propositional symbol $p\in\Pi_\P$, the following chain:
\begin{eqnarray*}
\lefteqn{T_\P(J)(p)=[T_\P^1(J)(p),T_\P^2(J)(p)]=}\\[15pt]
	&=&\!\!\!\!\Big[\!\sup\{T_{\P_{\gamma^p_i}}^1\!(J)(p)\mid i\!\in\!\{1,\dots,m_p\}\},\sup\{T_{\P_{\gamma^p_i}}^2(J)(p)\mid i\!\in\!\{1,\dots,m_p\}\}\Big]\\[15pt]
	&=&\!\!\!\!\Big[\!\max\{T_{\P_{\gamma^p_i}}^1\!(J)(p)\mid i\!\in\!\{1,\dots,m_p\}\},\max\{T_{\P_{\gamma^p_i}}^2\!(J)(p)\mid i\!\in\!\{1,\dots,m_p\}\}\Big]
\end{eqnarray*}
Now, we will work individually with each component of the immediate consequence operator. Given two $\mathcal{C}([0,1])$-interpretations $J_1$ and $J_2$, we will build two programs $\P^1$ and $\P^2$ from the rules of $\P$ such that, for each propositional symbol $p\in\Pi_\P$, the following inequalities are satisfied:
$$|T_{\P}^1(J_1)(p) - T_{\P}^1(J_2)(p)| \leq |T_{\P^1}^1(J_1)(p) - T_{\P^1}^1(J_2)(p)|$$
$$|T_{\P}^2(J_1)(p) - T_{\P}^2(J_2)(p)| \leq |T_{\P^2}^2(J_1)(p) - T_{\P^2}^2(J_2)(p)|$$

Specifically, for each propositional symbol $p\in\Pi_\P$, we will add a rule with head $p$ from the original program to $\P^1$ and a rule with head $p$, which can be  the same or other different rule, from the original program to $\P^2$.

Let us begin then with the computation of the program $\P^1$, which we suppose empty by default, this is, without any rule. Let $p\in\Pi_\P$. As $\P$ has a finite number of rules, there exist $\gamma_1,\gamma_2\in\Gamma$ such that
\begin{eqnarray}
 T_{\P}^1(J_1)(p)&=&\max\{T_{\P_{\gamma^p_i}}^1(J_1)(p)\mid i\in\{1,\dots,m_p\}\}=T_{\P_{\gamma_1}}^1(J_1)(p)\label{eqn:Tp_max1}\\
T_{\P}^1(J_2)(p)&=&\max\{T_{\P_{\gamma^p_i}}^2(J_2)(p)\mid i\in\{1,\dots,m_p\}\} =T_{\P_{\gamma_2}}^1(J_2)(p)\label{eqn:Tp_max2}
\end{eqnarray}

Suppose that $T_{\P_{\gamma_2}}^1(J_2)(p)\leq T_{\P_{\gamma_1}}^1(J_1)(p)$. Then, by Equation~(\ref{eqn:Tp_max2}), the following chain of inequalities holds:

$$T_{\P_{\gamma_1}}^1(J_2)(p)\leq T_{\P_{\gamma_2}}^1(J_2)(p)\leq T_{\P_{\gamma_1}}^1(J_1)(p)$$

Therefore
$$|T_{\P_{\gamma_1}}^1(J_1)(p) - T_{\P_{\gamma_2}}^1(J_2)(p)| \leq |T_{\P_{\gamma_1}}^1(J_1)(p) - T_{\P_{\gamma_1}}^1(J_2)(p)|$$

Consequently, we add the rule of the program $\P_{\gamma_1}$ with head $p$ to the program $\P^1$.

On the other hand, if $T_{\P_{\gamma_1}}^1(J_1)(p)\leq T_{\P_{\gamma_2}}^1(J_2)(p)$, taking into account Equation~(\ref{eqn:Tp_max1}), we obtain the following chain:

$$T_{\P_{\gamma_2}}^1(J_1)(p)\leq T_{\P_{\gamma_1}}^1(J_1)(p)\leq T_{\P_{\gamma_2}}^1(J_2)(p)$$

Thus
$$|T_{\P_{\gamma_1}}^1(J_1)(p) - T_{\P_{\gamma_2}}^1(J_2)(p)| \leq |T_{\P_{\gamma_2}}^1(J_1)(p) - T_{\P_{\gamma_2}}^1(J_2)(p)|$$

In this case, we add the rule of the program $\P_{\gamma_2}$ with head $p$ to $\P^1$.

Progressing with an analogous reasoning with the other propositional symbols, we obtain a program $\P^1$ such that for each propositional symbol $p\in \Pi_\P$ at most one rule with head $p$ appears in $\P^1$ and
\begin{eqnarray}\label{des:TP_P1}
|T_{\P}^1(J_1)(p) - T_{\P}^1(J_2)(p)| &\leq& |T_{\P^1}^1(J_1)(p) - T_{\P^1}^1(J_2)(p)|
\end{eqnarray}
Likewise, it is analogously obtained a program $\P^2$ such that for each propositional symbol $p\in\Pi_\P$ we obtain that
\begin{eqnarray}\label{des:TP_P2}
|T_{\P}^2(J_1)(p) - T_{\P}^2(J_2)(p)| &\leq& |T_{\P^2}^2(J_1)(p) - T_{\P^2}^2(J_2)(p)|
\end{eqnarray}

Since Equations \eqref{des:TP_P1} and \eqref{des:TP_P2} hold, for all $p\in \Pi_\P$, we obtain that:
\begin{eqnarray*}
||T_{\P}^1(J_1) - T_{\P}^1(J_2)||_\infty&\leq& ||T_{\P^1}^1(J_1) - T_{\P^1}^1(J_2)||_\infty\\
||T_{\P}^2(J_1)- T_{\P}^2(J_2)||_\infty&\leq& ||T_{\P^2}^2(J_1) - T_{\P^2}^2(J_2)||_\infty
\end{eqnarray*}

Moreover, by the definition of $\overline{T_{\P}}$, the following inequalities are satisfied:
\begin{eqnarray*}
||\overline{T_{\P}}^1(\overline{J_1}) - \overline{T_{\P}}^1(\overline{J_2})||_\infty&\leq& ||\overline{T_{\P^1}}^1(\overline{J_1}) - \overline{T_{\P^1}}^1(\overline{J_2})||_\infty\\
||\overline{T_{\P}}^2(\overline{J_1}) - \overline{T_{\P}}^2(\overline{J_2})||_\infty&\leq& ||\overline{T_{\P^2}}^2(\overline{J_1}) - \overline{T_{\P^2}}^2(\overline{J_2})||_\infty
\end{eqnarray*}

These inequalities provide the  contractivity of $\overline{T_\P}$ in $A$ as we show next. By Equation~(\ref{norma_por_componentes}) and by the definition of $\P^1$ and $\P^2$, we obtain that
{\small
\begin{eqnarray*}
	||\overline{T_\P}(\overline{J_1})-\overline{T_\P}(\overline{J_2})||_\infty \!\!&=&\!\!\!
	\max\!\Big\{||\overline{T_\P}^1(\overline{J_1})-\overline{T_\P}^1(\overline{J_2})||_\infty,||\overline{T_\P}^2(\overline{J_1})-\overline{T_\P}^2(\overline{J_2})||_\infty\!\Big\}\\
	&\leq&\!\!\!\max\Big\{||\overline{T_{\P^1}}^1(\overline{J_1}) - \overline{T_{\P^1}}^1(\overline{J_2})||_\infty\,||\overline{T_{\P^2}}^2(\overline{J_1}) - \overline{T_{\P^2}}^2(\overline{J_2})||_\infty\Big\}
\end{eqnarray*}
}
Moreover, since $\P^1$ and $\P^2$ are programs such that each propositional symbol $p\in\Pi_\P$ has at most one rule in $\P^1$ and one rule in $\P^2$ with head $p$, both programs are in the situation of the base case. As a consequence, $\overline{T_{\P^1}}^1$ and $\overline{T_{\P^2}}^2$ are both contractive mappings in $A$ with Lipschitz constants $\lambda^1$ and $\lambda^2$, respectively. Taking into account the first component of $\overline{T_{\P^1}}$ and the second of $\overline{T_{\P^2}}$, we deduce that
{\small
\begin{eqnarray*}
||\overline{T_\P}(\overline{J_1})-\overline{T_\P}(\overline{J_2})||_\infty
	&\leq&\!\!\!\max\Big\{||\overline{T_{\P^1}}^1(\overline{J_1}) - \overline{T_{\P^1}}^1(\overline{J_2})||_\infty\,||\overline{T_{\P^2}}^2(\overline{J_1}) - \overline{T_{\P^2}}^2(\overline{J_2})||_\infty\Big\}\\
	&\leq&\max\Big\{||\overline{J_1}-\overline{J_2}||_\infty\cdot\lambda^1,||\overline{J_1}-\overline{J_2}||_\infty\cdot\lambda^2\Big\}\\[5pt]
	&\leq&\max\Big\{||\overline{J_1}-\overline{J_2}||_\infty,||\overline{J_1}-\overline{J_2}||_\infty\Big\}\cdot\max\{\lambda^1,\lambda^2\}\\[5pt]
	&=& ||\overline{J_1}-\overline{J_2}||_\infty\cdot\max\{\lambda^1,\lambda^2\}
\end{eqnarray*}
}
Therefore, $\overline{T_\P}$ is a contractive mapping in $A$ whose Lipschitz constant is equal to $\max\{\lambda^1,\lambda^2\}=\lambda^2$. Applying Banach fix-point theorem, we obtain that $\overline{T_\P}$ has a unique fix-point in $A$.
		
Now, we will prove that $T_\P$ has a unique fix-point in $\mathcal{C}([0,1])^n$. For all $\mathcal{C}([0,1])$-interpretation $I$ being a fix-point of $T_\P$, we will demonstrate that $\overline {I}\in A$ and $\overline {I}$ is a fix-point of  $\overline{T_\P}$.

We suppose that $I$ is a fix-point of $T_\P$, that is, $I$  is a $\mathcal{C}([0,1])$-interpretation such that $T_\P(I)=I$. On the one hand, considering the inclusion mapping $\iota$, we can ensure that $\overline{T_\P}(\overline {I})= \iota (T_\P(I))=\iota (I)= \overline {I}$. Therefore, $\overline{I}$ is a fix-point of $\overline{T_\P}$. On the other hand, we will evince that $\overline {I}\in A$. Clearly, the inequality $T_\P(I)(p)\leq\max\{[\vartheta^1, \vartheta^2]\!\mid\!\langle p\leftarrow_{\beta_{w}\delta_{w}}^{\alpha_{w}\gamma_{w}} \mathcal{B} ;[\vartheta^1, \vartheta^2]\rangle\in\P\} $ holds, for all $\mathcal{C}([0,1])$-interpretation $I$. As a consequence, $I=T_\P(I)\sqsubseteq I_\vartheta$ and so, by definition of set $A$, we obtain that $\overline {I}\in A$.

Finally, by Proposition~\ref{prop:mfp} and Theorem~\ref{thm:existencia}, we can conclude that the only fix-point of $T_\P$ is actually the only stable model of the program.
\qed
\end{proof}

The following example shows a simple MANLP which satisfies the hypothesis of Theorem~\ref{th:uni}.

\begin{example}\label{ex:ultimo} Given the set of propositional symbols $\Pi_\P=\{p,q,s,t\}$ and the multi-adjoint normal lattice $(\mathcal{C}([0,1]),\leq,\leftarrow_{11}^{11}, \adjoint_{11}^{11},\leftarrow_{12}^{23}, \adjoint_{12}^{23},\leftarrow_{11}^{22}, \adjoint_{11}^{22}, \neg )$, we will define the multi-adjoint normal logic program $\P$ as follows:
\begin{equation*}
	\begin{array}{l}
	r_1:\ \langle p\leftarrow_{11}^{11} \ \neg q\ ;\ [0.7,0.8]\rangle\\
	r_2:\ \langle s\leftarrow_{12}^{23} \ p\ ;\ [0.4,0.5]\rangle\\
	r_3:\ \langle p\leftarrow_{11}^{22} \ s * \neg t\ ;\ [0.5,0.6]\rangle\\
	r_4:\ \langle q\leftarrow_{11}^{22} \ t * \neg p\ ;\ [0.7,0.9]\rangle
		\end{array}
	\end{equation*}
In order to apply Theorem~\ref{th:uni} to the program $\P$, we need to verify that the inequality
{\footnotesize
$$
\sum_{j=1}^{h}(\vartheta^2)^{\beta_w}\cdot \delta_w\cdot (\vartheta^2_{q_j})^{\delta_w-1}\cdot\Big(\vartheta^2_{q_1}\!\cdots \vartheta^2_{q_{j-1}}\! \cdot\, \vartheta^2_{q_{j+1}}\cdots\vartheta^2_{q_h}\Big)^{\delta_w}\!\!+(\vartheta^2)^{\beta_w}\cdot \delta_w\cdot (k-h)(\vartheta^2_{q_1}\!\cdots \vartheta^2_{q_h})^{\delta_w}\!< \!1
$$}
holds for every rule $\langle p\leftarrow_{\beta_{w}\delta_{w}}^{\alpha_{w}\gamma_{w}} q_1*\cdots*q_h*\neg q_{h+1}*\cdots*\neg q_k; [\vartheta^1,  \vartheta^2]\rangle\in\P$.

Concerning rule $r_1$, we obtain that
$$0.9< 1$$
For $r_2$, it holds that
$$0.5\cdot 2\cdot 0.9=0.9 < 1$$
For $r_3$ we obtain: $0.6+0.5\cdot 0.6=0.9< 1$.
Finally, for $r_3$ we have
$$
0.9+0.9\cdot 0=0.9< 1
$$
Therefore, the program $\P$ satisfies the hypothesis of the uniqueness theorem (Theorem~\ref{th:uni}). As a consequence, we can ensure that it has a unique stable model.

Moreover, this stable model can be computed by iterating $T_\P$ under the minimum interpretation $I_\bot$. The following table shows the iterations of the immediate consequence operator, taking as first the entry interpretation constantly $\bot$, which is denoted as $I_\bot$.

\medskip
\begin{center}
\begin{tabular}{|c||c|c|c|c|}
\hline
  & $p$ & $q$ & $s$ & $t$\\[1ex]
\hline
$I_\bot$ & [0,0] & [0,0] & [0,0] & [0,0]\\
\hline
$T_\P(I_\bot)$ & [0.7,0.9] & [0,0] & [0,0] &[0,0]\\
\hline
$T_\P^2(I_\bot)$ & [0.7,0.9] & [0,0] & [0.05488,0.405] &[0,0]\\
\hline
$T_\P^3(I_\bot)$ & [0.7,0.9] & [0,0] & [0.05488,0.405] &[0,0]\\
\hline
\end{tabular}
\end{center}
\medskip

Thus, $T_\P^2(I_\bot)$ is the unique stable model of $\P$.\qed
\end{example}

The previous example has also shown how the unique stable model can be computed. We only need to iterate the operator $T_\P$ under $I_\bot$ and the obtained fix-point will be the stable model. As a consequence,  when the hypotheses of the unicity  theorem hold, the computational complexity for computing the unique stable model is the same as in the case of positive logic programs. Hence, in this case, the use of negation operators does not  increase the  complexity for obtaining consequences from the knowledge system represented by the multi-adjoint normal logic program.

\section{Conclusions and future work}\label{sec:conclusion}

This paper has considered the philosophy of the multi-adjoint paradigm  to introduce the syntax and semantics of a new and more flexible normal logic programming framework.  We have proven that the stable models in multi-adjoint normal  logic programs satisfy the more important properties of the classical and residuated case. Moreover,  we have proven the existence of   stable models  of logic programs defined on a multi-adjoint normal lattice whose carrier is a convex compact set and the operators are continuous. Furthermore, a special kind of multi-adjoint normal logic program on the subinterval lattice $(\mathcal{C}([0,1]),\leq)$ has been introduced in which the uniqueness of stable models is ensured. Moreover, this stable model can easily be computed by iterating the immediate consequence operator from the minimum interpretation.

The  introduced results  on the existence and uniqueness of stable models can straightforwardly be applied to different useful frameworks, such as    monotonic and residuated logic programming~\cite{DP01:ecsqaru,DP-jelia},  fuzzy logic programming~\cite{vojtas-fss} and  possibilistic logic programming~\cite{dlp:1994}, in which a negation operator is considered in the language. In addition, we can use the introduced results in fuzzy answer set programming when the notion of x-consistency is included in our logical framework~\cite{Van_Nieuwenborgh_2007}.
These results may also be applied  to  generalized annotated logic programs~\cite{gaps}, when a negation operator is used. Taking into consideration the relation given in~\cite{klv} to the fuzzy logic programming, we can define a new generalized annotated logic programming in which a negation operator (decreasing with respect to the  ordering considered for the $T_\P$ operator) can be used and, as a consequence, a more flexible annotated logic is obtained.  Moreover, the proposed  framework in this paper will be also compared with the paraconsistent logic programming~\cite{damasio2002,damasio2005}, the disjunctive logic programming~\cite{bauters2014,MINKER1990,nieves2011,stracciaOD09} and the sorted logic programming~\cite{Baral2009,jal05}.
These relations are more complex and they will be studied in depth  in the future.
In addition, we will apply the obtained results to other logics with negation operators, such as  Possibilistic Defeasible Logic Programming (PDeLP)~\cite{Alsinet2008}.

The representation of knowledge from databases containing uncertainty and inconsistent information is an important task. Different authors~\cite{Gelfond1990,Kowalski1991,Pereira92,Wagner1991} have highlighted that, in order to handle this goal in a suitable way,   it is necessary to distinguish what can be proved to be false from what is false   because it cannot be proved true, which is called false by default.  This difference  can be obtained by the use of an explicit negation operator in normal logic programs. This philosophy is used by the  paraconsistent logic programming framework given in~\cite{damasio2002,damasio2005}, where the explicit and default negations have been considered and  related by  the coherence principle. Therefore, the relation between paraconsistent logic programming~\cite{damasio2002,damasio2005} and multi-adjoint normal logic programming will be also important for studying the notion of inconsistency in MANLPs.

Moreover, we will also adapt the definitions of coherence and inconsistency given in~\cite{NMadrid3,madrid:2011} to the multi-adjoint framework, which are focused on the  detection of plausible stable models. In particular,  we will  inspect   measuring inconsistency in fuzzy answer set semantics for MANLPs. Furthermore, real-life applications will be considered in which the introduced flexible multi-adjoint framework will be applied.

\end{document}